\documentclass{aa}
\usepackage[varg]{txfonts}
\usepackage[utf8]{inputenc}
\usepackage{wasysym}
\usepackage{natbib}
\usepackage{graphicx}
\usepackage{gensymb}
\usepackage{textcomp}
\usepackage{hyperref}
\usepackage{booktabs}
\usepackage{enumitem}
\usepackage{amsmath}

\bibpunct{(}{)}{;}{a}{}{,}

\begin{document}

\defcitealias{koda2012imf}{K12}

\title{Constraining the top-light initial mass function in the extended ultraviolet disk of M83}

\author{R.~P.~V.~Rautio \inst{\ref{inst1}}
  \and A.~E.~Watkins \inst{\ref{inst2}}
  \and H.~Salo \inst{\ref{inst1}}
  \and A.~Venhola \inst{\ref{inst1}}
  \and J.~H.~Knapen \inst{\ref{inst3},\ref{inst4}}
  \and S.~Comer\'on \inst{\ref{inst4},\ref{inst3}}}

\institute{Space Physics and Astronomy research unit, University of Oulu, 90014 Oulu, Finland \\\email{riku.rautio93@gmail.com} \label{inst1}
  \and Centre of Astrophysics Research, School of Physics, Astronomy and Mathematics, University of Hertfordshire, Hatfield AL10 9AB, UK \label{inst2}
  \and Instituto de Astrof\'isica de Canarias, 38205 La Laguna, Tenerife, Spain \label{inst3}
  \and Departamento de Astrof\'isica, Universidad de La Laguna, 38200 La Laguna, Tenerife, Spain \label{inst4}}

\abstract{The universality or non-universality of the initial mass function (IMF) has significant implications for determining star formation rates and star formation histories from photometric properties of stellar populations.}
         {We reexamine whether the IMF is deficient in high-mass stars (top-light) in the low-density environment of the outer disk of M83 and constrain the shape of the IMF therein.}
         {Using archival \emph{Galaxy Evolution Explorer} (\emph{GALEX}) far ultraviolet (FUV) and near ultraviolet (NUV) data and new deep OmegaCAM narrowband H$\alpha$ imaging, we constructed a catalog of FUV-selected objects in the outer disk of M83. We counted H$\alpha$-bright clusters and clusters that are blue in FUV$-$NUV in the catalog, measured the maximum flux ratio $F_{\textrm{H}\alpha}/f_{\lambda \textrm{FUV}}$ among the clusters, and measured the total flux ratio $\Sigma F_{\textrm{H}\alpha}/\Sigma f_{\lambda \textrm{FUV}}$ over the catalog. We then compared these measurements to predictions from stellar population synthesis models made with a standard Salpeter IMF, truncated IMFs, and steep IMFs. We also investigated the effect of varying the assumed internal extinction on our results.}
         {We are not able to reproduce our observations with models using the standard Salpeter IMF or the truncated IMFs. It is only when assuming an average internal extinction of $0.10 < A_{\textrm{V}} < 0.15$ in the outer disk stellar clusters that models with steep IMFs ($\alpha > 3.1$) simultaneously reproduce the observed cluster counts, the maximum observed $F_{\textrm{H}\alpha}/f_{\lambda \textrm{FUV}}$, and the observed $\Sigma F_{\textrm{H}\alpha}/\Sigma f_{\lambda \textrm{FUV}}$.}
         {Our results support a non-universal IMF that is deficient in high-mass stars in low-density environments.}


\maketitle

\section{Introduction}
Star formation in the bright central regions of massive galaxies is a well-studied phenomenon. The past evolution of stellar populations in these regions can be satisfactorily characterized by standard initial mass functions (IMFs) and star formation histories  (SFHs; e.g., \citealt{kroupa2002imf}). However, star formation is not confined only to these dense, matter-rich regions of galaxies. On the contrary, active star formation has also been found in low surface brightness (LSB) dwarf galaxies (e.g., \citealt{sargent1970dwarfs, prole2019udg}) as well as in the diffuse outskirts of massive spirals (e.g., \citealt{gildepaz2005xuv}). The LSB regime where this diffuse star formation is found requires sensitive instrumentation and dedicated observation strategies, and as such, it is much less studied than the bright central regions of nearby galaxies.

A central question regarding star formation in different environments is the universality or non-universality of the IMF. This has significant implications, as knowing the shape of the IMF is of critical importance in determining star formation rates (SFRs) and SFHs from the photometric properties of stellar populations. The universality of the IMF has been supported by many observations (e.g., \citealt{kroupa2002imf, bastian2010imf, koda2012imf, watkins2017imf}). However, it has also been suggested that the IMF may vary depending on environmental parameters, such as gas density. In particular, the IMF may be deficient in high-mass stars in low-density environments, either due to a lower upper mass cutoff \citep{krumholz2008sf} or a steeper slope \citep{elmegreen2004imf}. Intrinsically invariant IMFs may also appear top-light in integrated observations due to stochastic effects \citep{pflamm2008imf}. Surveys spanning large ranges of luminosities and galaxy masses have indeed revealed trends favoring top-light IMFs in faint, diffuse, and low star-formation activity galaxies \citep{hoversten2008imf, meurer2009imf, lee2009hafuv}, while an opposite trend favoring a top-heavy IMF has been found for starburst galaxies \citep{gunawardhana2011gama}.

\begin{figure*}[ht]
  \centering
  \includegraphics[width=\textwidth]{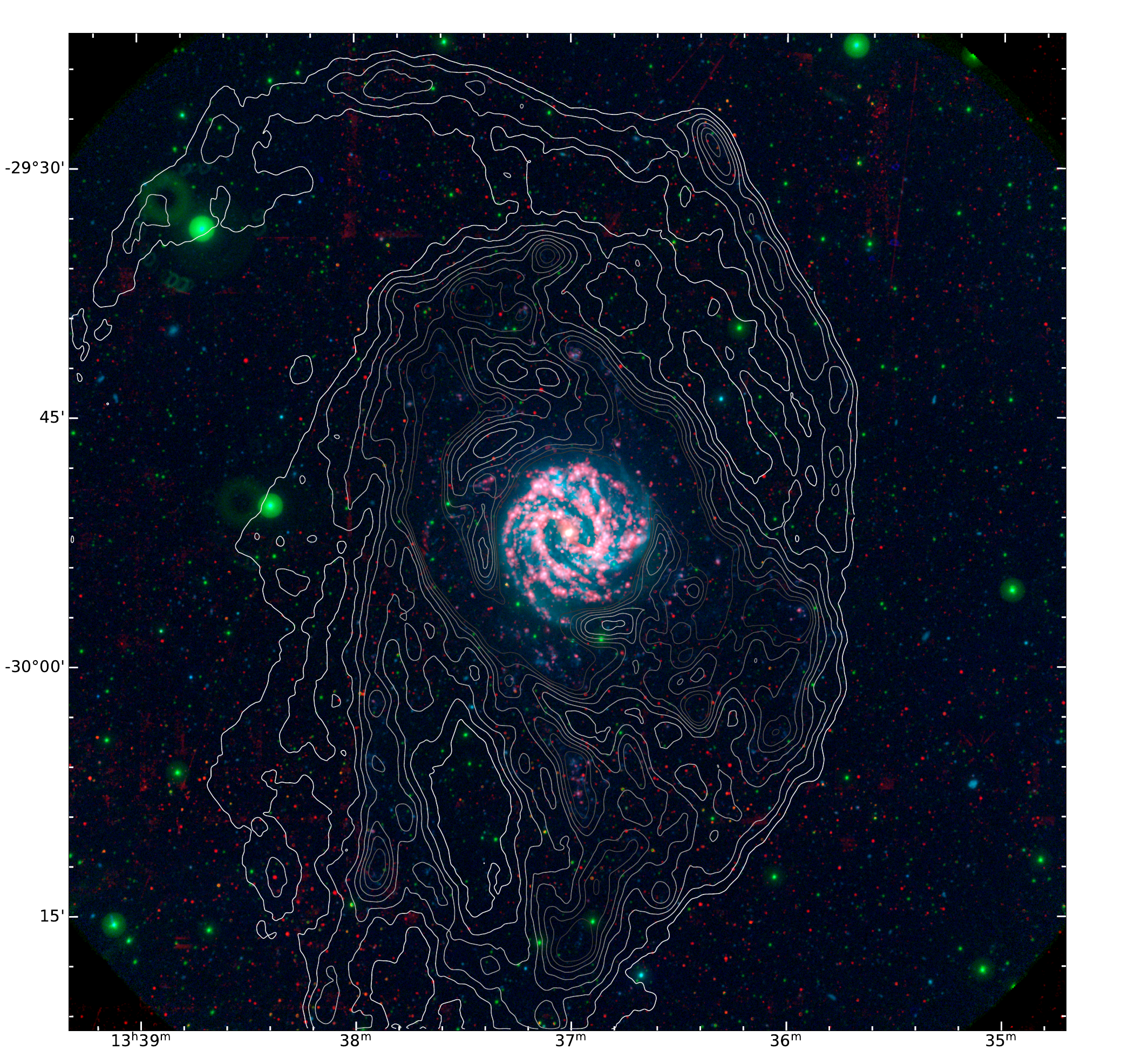}
  \caption{False color image of M83 and its XUV disk. Blue corresponds to FUV from \emph{GALEX}, green to NUV from \emph{GALEX}, and red to H$\alpha$ emission from this work. The H$\alpha$ image has been background-subtracted (see Sect. \ref{sec:pcbg}), and due to that, much of the diffuse H$\alpha$ light in the central regions of the galaxy has been removed. The contours show H{~\sc i} column densities going from $5\times 10^{19}$ cm$^{-2}$ (white contour) up to $4\times 10^{20}$ cm$^{-2}$ (darkest contour) in $5\times 10^{19}$ cm$^{-2}$ intervals. The contour levels were chosen to highlight the outer parts of the H{~\sc i} disk without covering the FUV-bright and H$\alpha$-bright parts of the XUV disk. The H{~\sc i} data is from the Local Volume H{~\sc i} survey (LVHIS; \citealt{koribalski2018hi}). The image is $1^\circ \times 1^\circ$ in size.}
  \label{fig:m83}
\end{figure*}

The \emph{Galaxy Evolution Explorer (GALEX)} Nearby Galaxy Survey revealed extended ultraviolet disks (XUV disks) around many nearby spiral galaxies \citep{gildepaz2005xuv, thilker2005m83, thilker2005m31, thilker2007xuv}. These XUV disks are an excellent test bed to study star formation in low-density environments, as far ultraviolet (FUV) imaging of \emph{GALEX} traces star formation that has occurred within the last 100 Myr \citep{kennicutt1998sf}. The prototypical XUV disk galaxy is M83 (NGC~5236), a nearby ($D=4.5$ Mpc; \citealt{karachentsev2002d}) grand design spiral, observations of which over many wavelengths have revealed molecular gas \citep{koda2022molecular}, neutral gas \citep{bigiel2010hi, heald2016hi, koribalski2018hi}, ionized gas (\mbox{\citealt{koda2012imf}}; hereafter K12), and young stellar clusters \citep{thilker2005m83, dong2008ir, bruzzese2020imf} far outside the classical bright, star-forming disk. Figure \ref{fig:m83} shows M83 and its XUV disk in FUV, near ultraviolet (NUV), H$\alpha$, and H{~\sc i}. Star formation in the XUV disk of M83 has been extensively studied, yet the picture is still not complete, as evidence both for \citepalias{koda2012imf} and against \citep{bruzzese2020imf} a universal IMF in the outer disk has been presented.

Ultraviolet emission of galaxies mainly originates from massive O and B stars down to $3 M_{\odot}$, with some contribution from evolved stars, such as white dwarfs, and post-asymptotic giant branch stars (e.g., \citealt{hills1972ism, flores-fajardo2011holmes, rautio2022}). To further constrain the upper mass range of the IMF, it is useful to look at the H$\alpha$ emission of H{~\sc ii} regions, which are ionized by FUV Lyman continuum photons ($\lambda < 912$ Å) originating primarily from massive O stars ($>20 M_\odot$). One of the biggest challenges with UV and H$\alpha$ data is dust extinction, and it especially affects UV data (e.g., \citealt{pei1992dust}). Dust may also increase the observed emission, specifically in galaxy outskirts, by scattering UV photons and potentially even H$\alpha$ photons originating from the bright inner parts of the galaxy \citep{ferrara1996scatter, wood1999scatter, hodges2014edust, seon2014edust, shinn2015edust, hodges2016edust, jo2018fuv}.

Deriving the IMF from observations requires modeling the time evolution of the spectra of stellar clusters and choosing an appropriate SFH. As clusters are born, hot massive OB stars dominate the emission, but as these luminous blue stars die, the cluster changes in color and emission intensity. The evolution of H$\alpha$ emission also depends on the dynamics of the H{~\sc ii} region, which in turn is affected by stellar winds and supernovae that may push the gas hundreds of parsecs away from the ionizing stars or even completely destroy the H{~\sc ii} region \citep{franco2000hii, churchwell2006bubble, whitmore2011hii, hannon2019hii}.

In this work, we use archival \emph{GALEX} FUV and NUV imaging and new deep OmegaCAM H$\alpha$ narrowband imaging to study the star formation in the M83 XUV disk. Our H$\alpha$ imaging covers a $1\fdg4 \times 1\fdg4$ area around M83, roughly equal to the \emph{GALEX} field of view (FoV). These data may be used to constrain the upper mass range of the IMF, as H$\alpha$ traces the massive O stars, while \emph{GALEX} FUV and NUV are sensitive to both O and B star emission at energies below the Lyman limit. We utilize stellar population synthesis modeling to test whether the standard \citet{salpeter1955imf} IMF, an IMF with a lower upper mass cutoff (truncated), or an IMF with a higher power-law index (steep) explain our observations best.

A similar analysis of the IMF in the M83 outer disk was performed by \citetalias{koda2012imf} using \emph{GALEX} UV data and H$\alpha$ data taken with the Subaru Prime Focus Camera (Suprime-Cam) on the Subaru telescope. However, recent work on the M83 outer disk IMF by \citet{bruzzese2020imf} with opposite conclusions from \citetalias{koda2012imf} warrants a reexamination of the H$\alpha$ emission of the M83 outer disk. Our H$\alpha$ data is also deeper\footnote{Our 1$\sigma$ photometric limit with 5\arcsec\ diameter aperture corresponds to a luminosity of $2.1 \times 10^{34}$ erg s$^{-1}$ at a distance $d = 4.5$ Mpc, compared to $7.97\times10^{34}$ erg s$^{-1}$ reported by \citetalias{koda2012imf}.} than that of \citetalias{koda2012imf} and covers more of the M83 outer disk and the surrounding background, allowing better background source contamination statistics. We also expand the analysis to test a set of truncated IMFs and a set of steep IMFs as well as different values of internal extinction.

We describe the observations and data reduction in Sect. \ref{sec:obs}. In Sect. \ref{sec:cata}, we construct a catalog of FUV-selected objects in the outer disk of M83 and investigate the H$\alpha$ and FUV luminosity functions and H$\alpha$-to-FUV coincidence of these objects. In Sect. \ref{sec:models}, we use stellar population synthesis modeling to compare the standard Salpeter IMF and IMFs with different slopes and truncation masses, and we obtain predictions for the numbers of H$\alpha$-bright clusters and clusters that are blue in FUV$-$NUV in the outer disk, the maximum H$\alpha$-to-FUV flux ratio of the clusters, and the H$\alpha$-to-FUV total flux ratio over the outer disk. In Sect. \ref{sec:results}, we compare these predictions to values measured from our catalog and investigate the effects of internal extinction on our results. We discuss the implications of our results for the IMF in low-density environments, compare our results to other work, discuss the limitations of our catalog, and discuss the effects that SFH, evolved stars, and scattered light may have on our results in Sect. \ref{sec:disc}. Finally, in Sect. \ref{sec:sum}, we summarize the results of this work.

\section{Observations and data reduction}
\label{sec:obs}

\subsection{OmegaCAM observations}

We obtained deep narrowband H$\alpha$ imaging of M83 with the OmegaCAM wide-field imager on the 2.6 m VLT Survey Telescope (VST) at the European Southern Observatory (ESO) Paranal Observatory in Chile. The OmegaCAM is a 32-CCD 16k $\times$ 16k detector mosaic with a $1^\circ \times 1^\circ$ FoV, which is enough to cover the entire XUV disk of M83. The OmegaCAM has a pixel scale of 0\farcs213/px. Notably, one of the detectors (CCD-77) was nonoperational during our observations.

\begin{figure}
  \centering
  \includegraphics[width=0.5\textwidth]{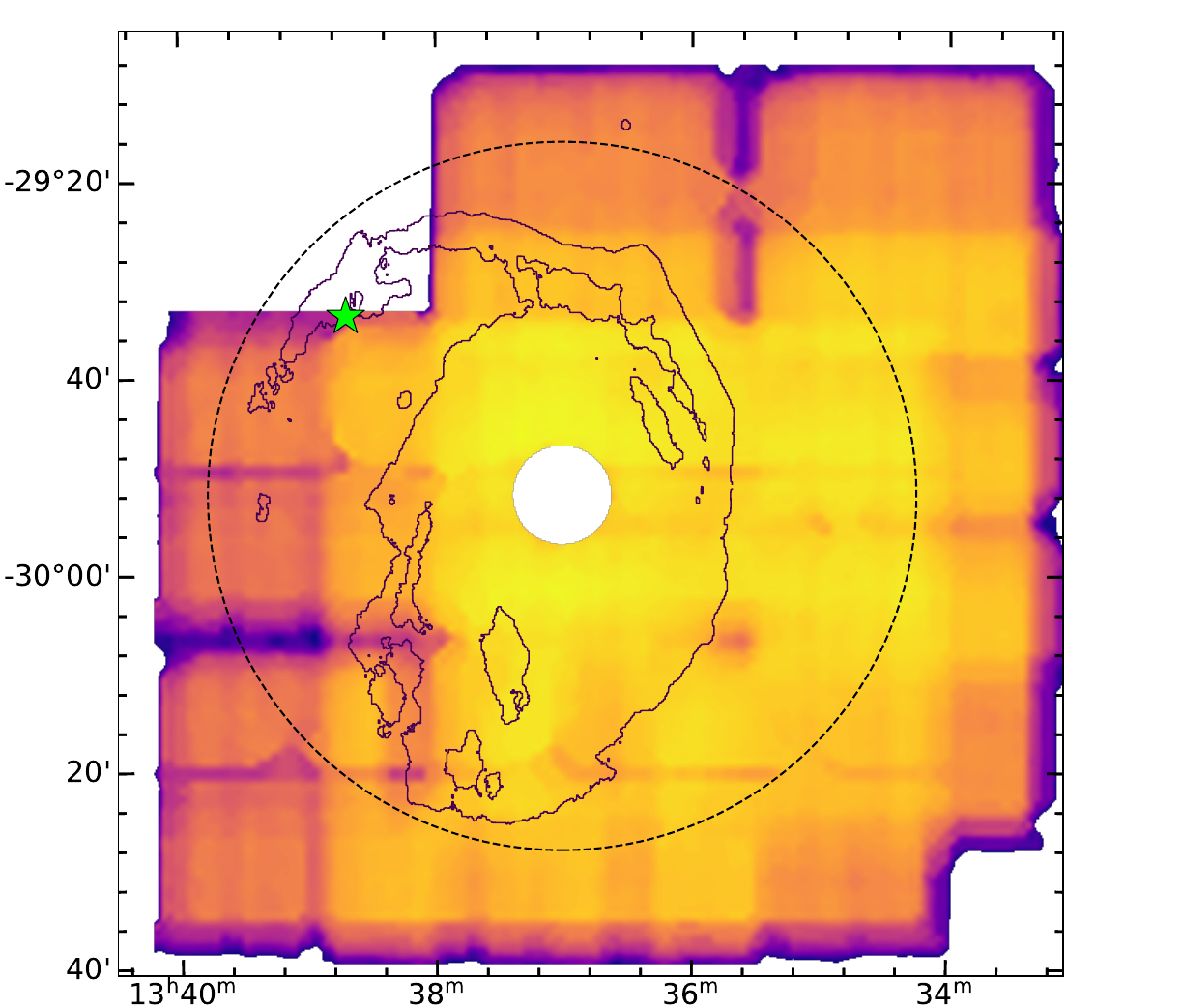}
  \caption{Sensitivity map of the H$\alpha$ imaging. Brighter colors correspond to a lower RMS and a higher exposure time. The inner disk within 5$\arcmin$ of the galaxy center is masked out. The black contour represents the $5 \times 10^{19}$ cm$^{-2}$ H{~\sc i} column density level. The black dashed circle shows the \emph{GALEX} FoV. The green star indicates the location of the fifth magnitude star HR 5128. The variation in background uncertainty is caused by the different exposure times in different areas due to the dithering pattern.}
  \label{fig:RMS}
\end{figure}

M83 was observed for 8 hours and 50 minutes in H$\alpha$ (filter NB\_659) and for 53 minutes in the Sloan \emph{r} (filter r\_SDSS) over 12 nights in December 2020 and January 2021. The H$\alpha$ filter was originally created and utilized by the VST Photometric H$\alpha$ Survey of the Southern Galactic Plane and Bulge (VPHAS+; \citealt{drew2014filter}). It is a segmented filter with a central wavelength of 659 nm ($\lambda = $ 658.6 nm for three segments and $\lambda = $659.3 for one segment) and a full width at half maximum (FWHM) of 10.5 nm. The interface of the four quadrants of the filter casts a cross-shaped shadow on the image plane. We used large dithering between each exposure to compensate for this vignetting and to capture enough sky for a robust background subtraction. The dithering pattern was chosen to avoid a fifth magnitude star (HR 5128) northeast of the galaxy. The effects of the dithering pattern to the depth in different parts of the image can be seen in Fig. \ref{fig:RMS}. A single pair of H$\alpha$ and \emph{r}-band exposures were centered on HR 5128 in order to obtain the extended point spread function (PSF) in our data (see Sect. \ref{sec:psf}), but these exposures were not used in the final coadds.

\subsection{Preliminary narrowband data reduction}
We began our data reduction by subtracting the row-wise median of the overscan region from each science and calibration image. Then nightly master bias images were created and subtracted from all science and flat-field images.

For our flat-field correction, we used an iterative process of constructing night-sky flat-fields from masked science images. To construct the preliminary flat-fields, we adopted the method of combining dome and twilight flat-fields that were used by the Kilo Degree Survey (KiDS; \citealt{dejong2015KiDS}) and by the Fornax Deep Survey (FDS; \citealt{venhola2018fds}). Dome and twilight master flat-fields were first created by median combination. Then, the high-frequency spatial Fourier modes were taken from the dome master flat and multiplied with the low-frequency spatial Fourier modes taken from the twilight master flat. This was done because the large-scale illumination of the twilight flat-field matches better with the observational situation than that of the dome flat-field, while the signal-to-noise of the dome flat-field is higher, which captures pixel-by-pixel variations better. This preliminary flat-field correction was then applied to the bias-subtracted science images, and astrophysical objects and reflection artifacts were masked out using NoiseChisel \citep{gnuastro,noisechisel}. Each CCD in the bias-subtracted science images was then normalized to a common image level, and the masks obtained using the preliminary flat-field correction were applied to them. These normalized and masked science images were then median combined to create a master sky flat-field. Using the same method as described above, this master sky flat-field was combined with the master dome flat-field to create the final flat-field. The bias-subtracted science images were then flattened with this final flat-field.

There were large variations in the sky background across the 12 nights of observations as well as within nights. M83 covered roughly two of the 32 OmegaCAM CCDs, and there were CCD-to-CCD differences in the sky pattern. To take the temporal sky variations into account while capturing the CCD-to-CCD sky differences, we constructed a sky background model for each hour-long observation block of four exposures. The models were constructed by masking out astronomical objects and artifacts from each exposure, scaling them to a common level, and median-combining the masked and scaled images. In order to reduce the pixel-to-pixel noise in the background models, we binned them with a 128 px $\times$ 128 px bin size, interpolated the flux across any remaining masked pixels, and applied a Gaussian smoothing using a kernel with a FWHM of half the bin size. This binned and smoothed background model was then scaled to the background level of each image in the observation block and subtracted out.

Preliminary coadds were then created by first registering each image to a common world coordinate system (WCS) with SCAMP \citep{bertin2006scamp}, using TPV projection and Gaia data release 3 \citep{gaia1, gaia2} as a reference catalog. We then combined the images with SWarp \citep{bertin2002swarp} using outlier-filtered mean (\texttt{CLIPPED}) stacking.

\subsection{Point spread function subtraction}
\label{sec:psf}
The preliminary coadds were used in constructing PSFs for both of the filters. We followed the recipe of \citet{infante-sainz2020psf} to construct a four-part extended PSF from stars of different magnitudes in our FoV. Faint unsaturated stars were used to construct the core of the PSF, while saturated, progressively brighter stars gave the outer parts. The three inner parts of the PSF (extending up to a radius of 2\arcsec, 8\arcsec, and 100\arcsec ,\ respectively) were built by stacking masked postage stamp images of stars (of G magnitude 13--12, 12--10, and 10--6, respectively) cut from the preliminary coadds. We used GNUastro scripts \citep{gnuastro,noisechisel} to build a segmentation map and construct the stacks. Before stacking, the postage stamp images were scaled to a common flux in annuli between 1\arcsec \ to 2\arcsec, 2\arcsec \ to 3\arcsec, and 4\arcsec \ to 5\arcsec, respectively for the three inner parts of the PSF. We then scaled and combined the three inner parts of the PSF along a 2\arcsec and an 8\arcsec radius, respectively. The PSFs in the H$\alpha$ filter and $r$-band are very similar, but not identical, as the H$\alpha$ PSF is slightly more centrally peaked (FWHM of 1\farcs00 vs. 1\farcs04).

\begin{figure}
  \centering
  \includegraphics[width=0.5\textwidth]{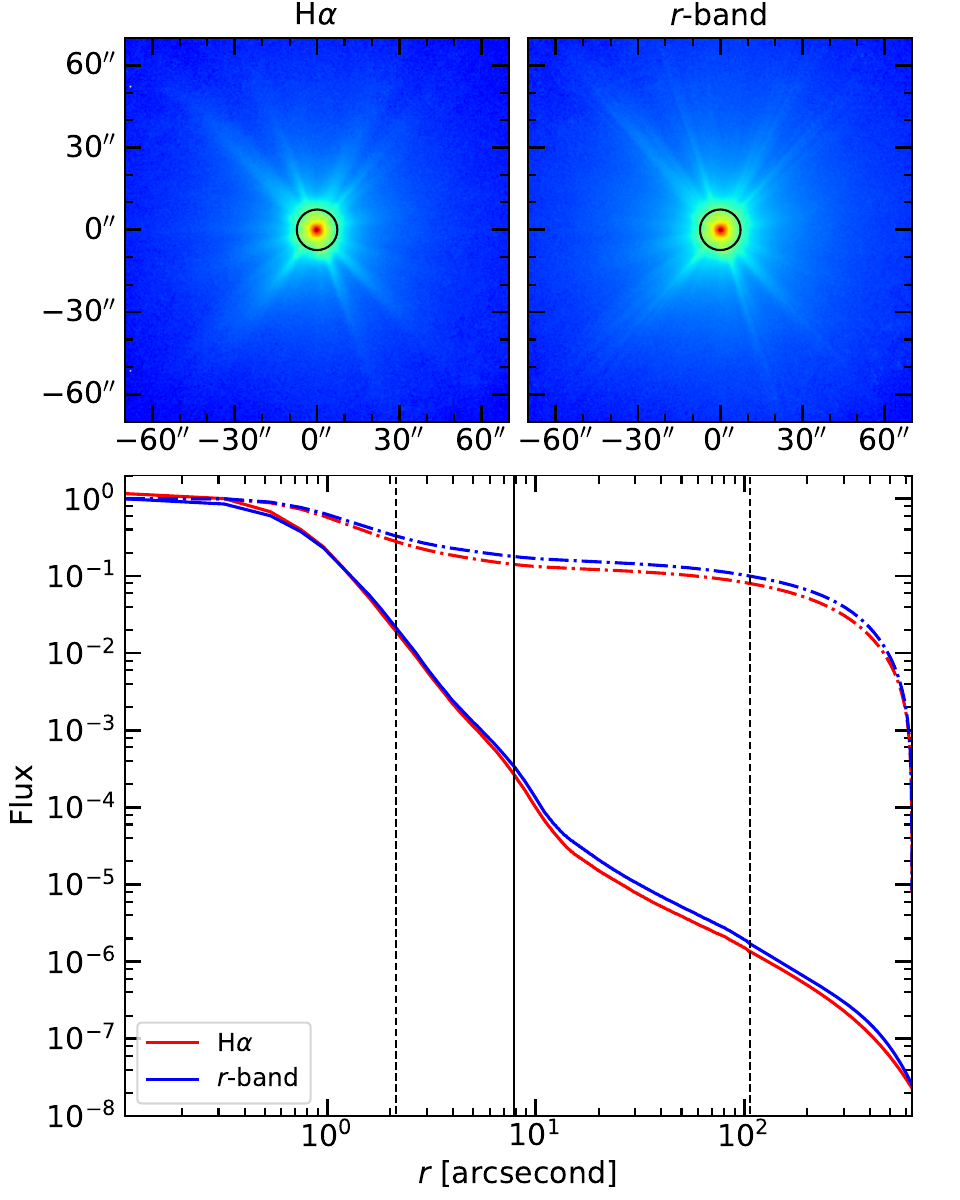}
  \caption{Point spread functions of our OmegaCAM data. \emph{Top-left panel:} Inner parts of the H$\alpha$ PSF. \emph{Top-right panel:} Inner parts of the \emph{r}-band PSF. \emph{Bottom panel:} Radial profile of the extended H$\alpha$ PSF (red curves) and \emph{r}-band PSF (blue curves). The \emph{r}-band PSF is normalized to its maximum value, and the H$\alpha$ PSF is normalized so that it has a total flux equal to the \emph{r}-band PSF. The dot-dashed curves show the fractional cumulative flux outside a given distance from the center. The black vertical lines correspond to the separations between the different parts of the PSFs (2\arcsec, 8\arcsec, and 100\arcsec; see text). The separation between the two middle parts (solid black line at 8\arcsec) is also shown in the top panels as a black circle.} 
  \label{fig:PSF}
\end{figure}

To model the extended wings of the PSF, we fit a fifth order polynomial to the radial profile of the fifth magnitude star HR 5128. Although the GNUastro scripts mask the worst reflections out of the postage stamp images, some residual reflection light still remained in the stacked PSF. To correct for this background of reflection light, we compared the radial profile of the stacked PSF to the radial profile of the fifth magnitude star, which due to landing in a CCD gap was much less affected by the reflections. We made a fit assuming that the radial profile of the stacked PSF is equal to the normalized radial profile of the fifth magnitude star plus a constant background. We then subtracted this fitted background value from the stacked PSF. Finally, we scaled and combined the stacked PSF to the modeled extended wings along a 100\arcsec radius. The combined PSFs extend up to 10\arcmin \ in radius. They are shown in Fig. \ref{fig:PSF}.

The extended PSFs were scaled to all stars brighter than $m_\mathrm{G} = 10$ using flux in an appropriate annulus chosen by hand and subtracted from the flattened science images. The sky background was then reestimated and subtracted from these star-subtracted images, following the process described above for each hourly block. Final coadds were generated from these star and background-subtracted images with SCAMP and SWarp using the same parameters as above.

\subsection{Continuum subtraction}
To subtract the stellar continuum from the H$\alpha$ image, the \emph{r}-band image must be scaled to be proportional to the continuum in the H$\alpha$ image. We estimated this scale factor by plotting the intensity of each pixel in the H$\alpha$ image against its intensity in the \emph{r}-band. Most of the pixels in the H$\alpha$ image lack line emission and have flux that is linearly proportional to the \emph{r}-band flux and thus appear as a straight line in the plot. Pixels containing H$\alpha$ line emission instead systematically appear above this line due to their excess flux compared to the stellar continuum. The slope of this line gives the factor to scale the \emph{r}-band image that is to be subtracted from the H$\alpha$ image \citep{knapen2005diff,knapen2006diff}. We adjusted this scale factor by eye so that in the continuum-subtracted image, no unphysical, extensive oversubtraction occurred, which would be visible as extended regions with negative flux, and residual foreground starlight was minimized. Following this, we obtained the continuum-subtracted H$\alpha$ image as H$\alpha = \textrm{NB\_659} - 0.83 r$.

\subsection{Photometric calibration and background subtraction}
\label{sec:pcbg}
To calibrate the flux in our image, we scaled it to the continuum-subtracted H$\alpha$ image of M83 from the Survey for Ionization in Neutral Gas Galaxies (SINGG; \citealt{meurer2006sing}). The flux was scaled so that the total flux within  a 4\arcmin -radius aperture centered on the galaxy is equal to the total flux within that aperture in the SINGG image. The aperture size was chosen to be roughly equal to the radial extent of the bright inner disk H{~\sc ii} regions.

Large background variations (larger than 2$\sigma$ of the background) remained in the final coadds after the mosaicking. These were caused by reflections of the many bright stars in the FoV and sky background fluctuations that we were unable to accurately model and subtract out due to the small number of exposures that we obtained for each observing night. We used SExtractor \citep{sextractor} to subtract the background from the continuum-subtracted H$\alpha$ image with a 200 px $\times$ 200 px ($42\farcs6 \times 42\farcs6$) background mesh size. As the size of the background mesh is smaller than the size of M83 in our data, this background subtraction also removes all of the smooth diffuse light at the scale of the galaxy. As our analysis focuses on H{~\sc ii} regions that are of much smaller size, this destruction of diffuse light does not affect it.

\subsection{Archival \emph{GALEX} data}
\label{sec:galex}
We retrieved archival \emph{GALEX} FUV and NUV imaging from the Mikulski Archive for Space Telescopes (MAST). The \emph{GALEX} FUV channel covers the wavelengths from 1344 to 1786 Å with an effective wavelength of 1538.6 Å, while the NUV channel covers the wavelengths from 1771 to 2831 Å with an effective wavelength of 2315.7 Å. The detector has a pixel scale of 1\farcs5/px and an effective angular resolution of ~5\arcsec. The \emph{GALEX} observations of M83 were carried out by \citet{bigiel2010hi} and are discussed therein. To restrict our analysis only to bright clusters, we used SExtractor to subtract the background and any galaxy scale diffuse UV emission from the FUV and NUV images using a 100 px $\times$ 100 px ($150\arcsec \times 150\arcsec$)  background mesh size. A larger mesh size compared to the H$\alpha$ data was possible due to the comparative smoothness of the \emph{GALEX} background.

\subsection{Matching the H$\alpha$ and UV data}
The \emph{GALEX} FUV data has a PSF with an FWHM of 4\farcs2 \citep{morrissey2007galex}, while our OmegaCAM H$\alpha$ imaging has a PSF with FWHM of 1\farcs0. We convolved our H$\alpha$ image with a Gaussian kernel with a FWHM$_{\textrm{kernel}}^2$ = FWHM$_{\textrm{FUV}}^2$ $-$ FWHM$_{\textrm{H}\alpha}^2$, and we resampled it to the \emph{GALEX} grid to match the resolution and the pixel size between our UV and H$\alpha$ data. We masked out reflections and other artifacts from the H$\alpha$ image and applied the same masking to the \emph{GALEX} data. Our H$\alpha$ data covers the entirety of the \emph{GALEX} FoV except for a small section northeast of M83 around the fifth magnitude star HR 5128, where we only took one pair of H$\alpha$ and \emph{r}-band exposures to help determine the extended PSF of our data.

The resolution in the \emph{GALEX} NUV band is worse than in the FUV band (PSF FWHM of 5\farcs3; \citealt{morrissey2007galex}), but we performed no convolution to match them. While this may cause a slight underestimation in the NUV flux, as some of it may be spread outside our isophotal apertures due to PSF effects, we choose to ignore it because our analysis focuses on the FUV and H$\alpha$ data, and convolving everything to the NUV PSF would unnecessarily reduce the resolution of our data in these bands. The NUV data already shares the same \emph{GALEX} grid as the FUV data, so we did not perform any resampling for it either.

To quantify the sensitivity of our convolved and resampled H$\alpha$ image, we measured the median and standard deviation in 40 px $\times$ 40 px ($1 \arcmin \times 1 \arcmin$) background boxes across the image. In the deepest parts of the image near the galaxy, we found $\sigma_{\textrm{bg}} = 2.9 \times 10^{-18}$ erg cm$^{-2}$ s$^{-1}$, while in the least sampled parts of the outskirts, we found $\sigma_{\textrm{bg}} = 4.6 \times 10^{-18}$ erg cm$^{-2}$ s$^{-1}$. These values refer to $\sigma$ calculated per resampled $1\farcs5 \times 1\farcs5$ pixels. Large-scale background variations are characterized by the standard deviation of the median values of the background boxes. For this, we found $\sigma = 1.4 \times 10^{-18}$ erg cm$^{-2}$ s$^{-1}$, which is smaller than the intrinsic noise in our data, confirming that our background subtraction was successful.

\begin{figure*}[ht]
  \centering
  \includegraphics[width=\textwidth]{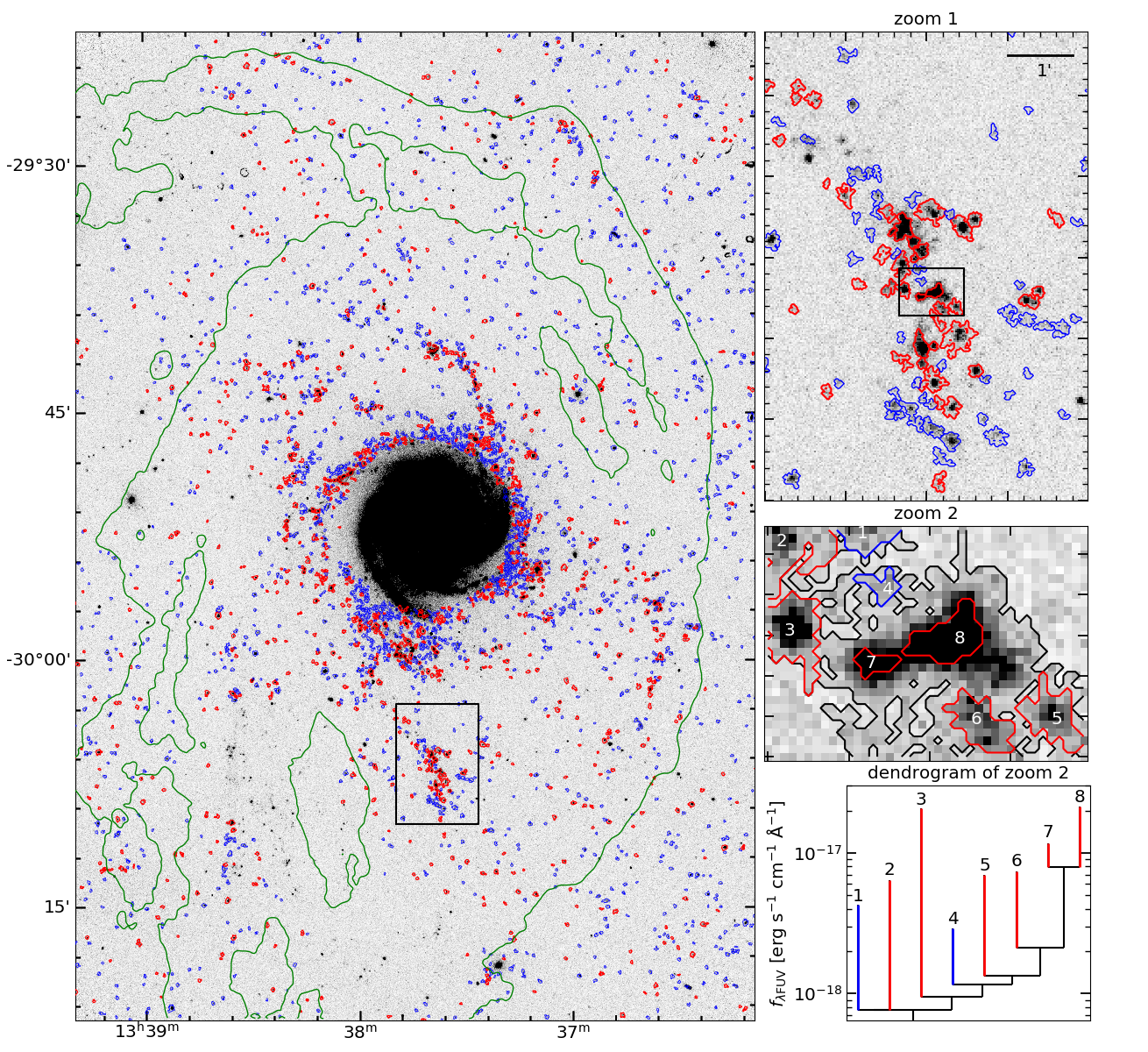}
  \caption{Dendrogram derived isophotal apertures of our sample objects. \emph{Left panel:} Apertures of our sample objects plotted over the \emph{GALEX} FUV image. Objects with 5$\sigma$ H$\alpha$ detection are outlined in red, while objects without 5$\sigma$ H$\alpha$ detection are outlined in blue. Objects flagged as belonging to the background, foreground, or artifacts, or with missing H$\alpha$ data, are not shown. The green contour shows the $5 \times 10^{19}$ cm$^{-2}$ H{~\sc i} column density level from LVHIS \citep{koribalski2018hi}. \emph{Top-right panel:} Zoom in of section outlined with a black rectangle in the left panel. A one arcminute scale bar is shown in the top-right corner. \emph{Middle-right panel:} Further zoom in of the section outlined with a black rectangle in the top-right panel. The lowest dendrogram level is outlined with a black contour. The objects within this zoomed in section are labeled with white numbers ranging from one to eight. \emph{Bottom-right panel:} Dendrogram of the data within the middle-right panel. The leaves corresponding with the objects in the middle-right panel are labeled with the same numbers as in the middle-right panel. The y-axis corresponds to the peak pixel flux within each leaf or the pixel flux in the saddle point between branches.}
  \label{fig:dendro}
\end{figure*}

\section{Catalog of FUV-selected objects}
\label{sec:cata}

\subsection{Sample selection}
We used the background-subtracted \emph{GALEX} FUV data to select young stellar clusters in the outer disk of M83. For this, we defined as the outer disk everything beyond 5\arcmin \ from the center of M83, where the H$\alpha$ flux drops significantly \citep{martin2001sf, thilker2005m83}, but within the radius of  the \emph{GALEX} FoV (36\arcmin). We found the clusters by computing dendrograms of the FUV data with the Astrodendro\footnote{\url{http://www.dendrograms.org/}} Python package and selecting the peaks or the ``leaves'' of the dendrogram as the stellar clusters. The Astrodendro algorithm works by constructing a hierarchical tree structure, a dendrogram, of a given dataset. A detailed description of constructing dendrograms can be found in \citet{goodman2009dendro}. We used 1$\sigma$ ($2.8 \times 10^{-19}$ erg s$^{-1}$ cm$^{-2}$ Å$^{-1}$; or 28 mag) as the minimum pixel value for the dendrograms, 3$\sigma$ as the minimum significance of a branch, and 10 pixels as the minimum size to consider a leaf to be an independent entity. This means that for Astrodendro to detect an object, the object must have a peak pixel value higher than 4$\sigma$ (the minimum significance plus the minimum pixel value) and nine additional connected pixels with values higher than 1$\sigma$. Astrodendro found 10\,405 objects fulfilling these criteria in the \emph{GALEX} FUV data of the M83 outer disk. We then summed the fluxes of the pixels identified by Astrodendro to belong to a single object and took that as the isophotal FUV flux of the stellar cluster. For structures consisting of only a single leaf, which is true for most of the objects in the \emph{GALEX} FUV image of the M83 outer disk, Astrodendro includes all connected pixels above the minimum value (1$\sigma$) as part of the structure. This ensured that for each object, we caught all the flux above the noise level of the data, and no flux would be lost to aperture effects. The isophotal apertures derived from the dendrogram leaves are shown in Fig. \ref{fig:dendro}.

To obtain the H$\alpha$ fluxes of the H{~\sc ii} regions associated with the young stellar clusters, we performed forced isophotal photometry on the H$\alpha$ image using the dendrogram leaves computed from the FUV data. To do this, we summed the H$\alpha$ fluxes of the pixels within a single dendrogram leaf footprint and took that as the flux of the associated H{~\sc ii} region. As H$\alpha$ emission does not always perfectly coincide with FUV emission (H{~\sc ii} region morphologies do not perfectly follow the distribution of the ionizing stars; \citealt{whitmore2011hii, hannon2019hii}), we may have missed a portion of the H$\alpha$ flux by using forced photometry. To test this, we performed a parallel analysis using only clusters where the H$\alpha$ emission was confirmed by eye to match the FUV emission perfectly. This did not affect our results, so we concluded that any H$\alpha$ flux missed by our forced photometry approach is insignificant compared to the natural variations in the flux. To obtain the NUV fluxes, we also performed forced photometry on the \emph{GALEX} NUV image using the same dendrogram leaf footprints.

\subsection{Extinction correction}
\label{sec:ext}
The emission from star clusters in the M83 outer disk is subject to internal extinction by the medium near the clusters ($A_V^{\textrm{int}}$) as well as to Galactic extinction by the interstellar medium (ISM) of the Milky Way ($A_V^{\textrm{MW}}$). For the Galactic extinction, we adopted a uniform value of $A_V^{\textrm{MW}} = 0.218$ mag \citep{schlegel1998ext}, as the ISM does not vary significantly within the approximate one square degree covered by the M83 outer disk. The internal extinction is harder to estimate since it may vary significantly from cluster to cluster \citep{gildepaz2007spec, bresolin2009spec}, and as such, adopting a single value increases the uncertainty in flux measurements.

We estimated the internal extinction for each object with H{~\sc i} data from the Local Volume H{~\sc i} survey (LVHIS; \citealt{koribalski2018hi}). We used the equation

\begin{equation}
  k = 9.21 \times 10^{20}(1 + R_V)(E_{B-V}/N_{\textrm{H{~\sc i}}}) \textrm{ cm}^{-2},
  \label{eq:ext}
\end{equation}

\noindent
where $k$ is the dimensionless dust-to-gas ratio, $R_V = A_V^\textrm{int}/E_{B-V}$ is the ratio of total-to-selective extinction, $E_{B-V}$ is the color excess, and $N_{\textrm{H{~\sc i}}}$ is the column density of neutral atomic hydrogen. As the low-metallicity and low-density environment of the outer disk is similar to that of dwarf galaxies, we use the Small Magellanic Cloud (SMC) values of $k=0.08$ and $R_V=2.93$ from \citet{pei1992dust}.

The internal extinction we obtain from the H{~\sc i} data is not very large: the average $A_V^\textrm{int} = 0.02$ mag for those objects that are within the H{~\sc i} disk (that have an H{~\sc i} detection in the LVHIS data). However, the resolution of the H{~\sc i} data is very poor, on the order of tens of arcseconds, and as such any local enhancements in the H{~\sc i} column density that would be expected near young stellar clusters and any dust associated with the stellar clusters themselves are not detected. Therefore, this estimate for internal extinction can be considered a lower limit. Nevertheless, assuming insignificant internal extinction is a reasonable first assumption, as a spectroscopic study by \citet{bresolin2009spec} found an average total extinction (including also the Galactic extinction) of only $A_{V} = 0.15$ mag for clusters with $R > R_{25}$ in the M83 outer disk, indicating $A_{V}^{\textrm{int}} \approx 0$.

On the other hand, \citet{gildepaz2007spec} report much higher total extinction in their spectroscopic study of stellar clusters in the outer disk of M83. The median extinction among their objects is $A_{V} = 0.62$ mag. There is also a large variation in the extinction between the clusters. Both \citet{bresolin2009spec} and \citet{gildepaz2007spec} report $A_{V} = 0$ mag for several clusters, while the highest extinction reported by \citet{bresolin2009spec} is $A_V = 1.16$ mag, whereas \citet{gildepaz2007spec} report $A_{V} = 1.55$ mag. The standard deviation among the sample of \citet{bresolin2009spec} is $A_V = 0.26$ mag, while among the sample of \citet{gildepaz2007spec}, it is $A_{V} = 0.49$ mag. Regarding the clusters present in both studies, \citet{gildepaz2007spec} report a nearly three times higher extinction. This variation in the extinction between clusters increases the uncertainty in our extinction correction. 

To account for the dependence of Galactic and internal extinction on different bands, we adopt the extinction curves of \citet{pei1992dust}, which give extinctions of $(A_{\textrm{FUV}}, A_{\textrm{NUV}}, A_{\textrm{H}\alpha}) = (2.58, 2.83, 0.81)A_{V}$ for the Galaxy and of $(4.18, 2.59, 0.79)A_{V}$ for the SMC. We adopted the extinction curve of the SMC for internal extinction due to the low metallicity of most objects in the outer disk of M83 ($\sim0.2 Z_{\odot}$; \citealt{gildepaz2007spec}). All the reported fluxes, magnitudes, colors, and flux ratios in this work were corrected for $A_V^{\textrm{MW}}$ and $A_V^\textrm{int}$. However, in Sect. \ref{sec:dext} we investigate how the results would change if $A_V^{\textrm{int}}$ were greater than estimated by Eq. \ref{eq:ext} from the LVHIS H{~\sc i} data.

\subsection{Rejecting background and foreground objects}
After correcting for Galactic and internal extinction, in order to reject background and foreground objects, data artifacts, and other spurious detections, we imposed several cuts to the cluster-candidate sample. To identify background galaxies, we searched for objects within the \emph{GALEX} FoV with measured radial velocities more than 500 km s$^{-1}$ greater than M83 from the NASA/IPAC Extragalactic Database (NED)\footnote{\url{https://ned.ipac.caltech.edu/}} and rejected them from the sample. To find and reject foreground stars, we used Gaia data release 3 \citep{gaia1, gaia2} to identify objects within the \emph{GALEX} FoV with parallax or proper motions larger than 3$\sigma$. For additional foreground star rejection, we looked at the continuum-subtracted H$\alpha$ image, where many foreground stars appear as rings with negative flux peaks due to oversubtraction. This oversubtraction happens due to a PSF mismatch between the \emph{r}-band and H$\alpha$ filters (see Fig. \ref{fig:PSF}). We rejected all objects containing pixels in the continuum-subtracted H$\alpha$ image with values smaller than $-5\sigma$ of the background. Finally, we chose objects with FUV$-$NUV colors and H$\alpha$-to-FUV flux ratios consistent with that of star clusters ($-1 < \textrm{FUV} - \textrm{NUV} < 2$ and log$(F_{\textrm{H}\alpha}/f_{\lambda\textrm{FUV}}) < 1.8$). This UV color range ensured that our sample contains clusters ranging from the youngest zero-age clusters (and even potential single O3 stars with FUV$-$NUV$\approx -0.6$, \citetalias{koda2012imf}) up to $\sim1$ Gyr old clusters \mbox{\citep{dong2008ir}} while rejecting red foreground stars and blue artifacts. The flux ratio limit rejects clusters with excess H$\alpha$ flux due to contamination by poorly subtracted stars or other artifacts in the H$\alpha$ image. After applying these cuts, we yielded a catalog of 4016 FUV-selected objects in the outer disk of M83.

\begin{figure}
  \centering
  \includegraphics[width=0.5\textwidth]{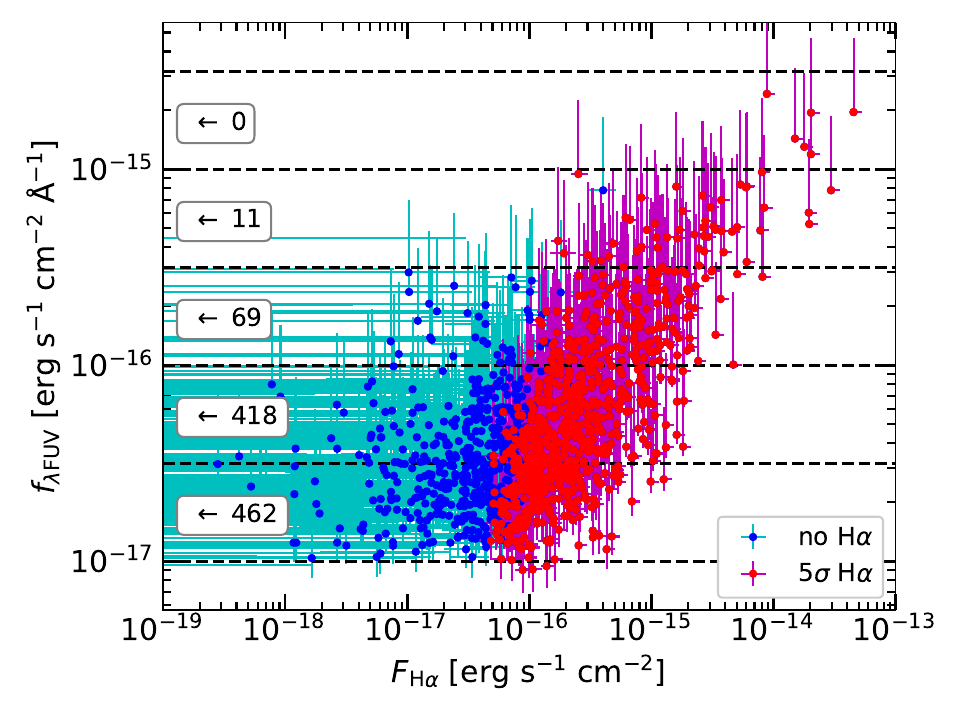}
  \caption{H$\alpha$ flux compared against FUV flux for the IN objects in our sample. The red dots are the objects with a 5$\sigma$ H$\alpha$ detection, and the blue dots are the objects without a 5$\sigma$ H$\alpha$ detection. The magenta and cyan lines give the error bars. The error bars are mainly due to uncertainty in extinction (see Sect. \ref{sec:error}). The dashed horizontal lines split the sample into logarithmic bins according to their FUV flux. The number of objects with negative H$\alpha$ flux for each bin is shown on the left.} 
  \label{fig:hafrac}
\end{figure}

Another method we used to reduce the contamination by background and foreground sources is statistical background subtraction. An H{~\sc i} contour ($N_{\textrm{H{~\sc i}}} = 1.5 \times 10^{20}$ cm$^{-2}$; \citealt{miller2009hi}) was used by \citetalias{koda2012imf} to separate their objects into IN and OUT samples in their study of stellar clusters in the outer disk of M83. They argued that objects outside the H{~\sc i} disk are most likely not associated with M83 and that they instead represent a background population present all across the FoV. By scaling the number of OUT objects with the IN/OUT area ratio and subtracting them from the IN objects, the background sources can be removed from the IN sample. We performed a similar statistical background subtraction but defined our IN area with $N_{\textrm{H{~\sc i}}} > 5 \times 10^{19}$ cm$^{-2}$ instead using the deep LVHIS data (see Fig. \ref{fig:m83}). There are 2249 objects in our IN area and 1767 objects in our OUT area ($N_\textrm{IN}/N_\textrm{OUT} = 1.27$), and the IN/OUT area ratio is $S_{\textrm{IN}}/S_{\textrm{OUT}} = 0.55$. The FUV and H$\alpha$ fluxes of our IN objects are plotted against each other in Fig. \ref{fig:hafrac}.

\begin{figure}
  \centering
  \includegraphics[width=0.5\textwidth]{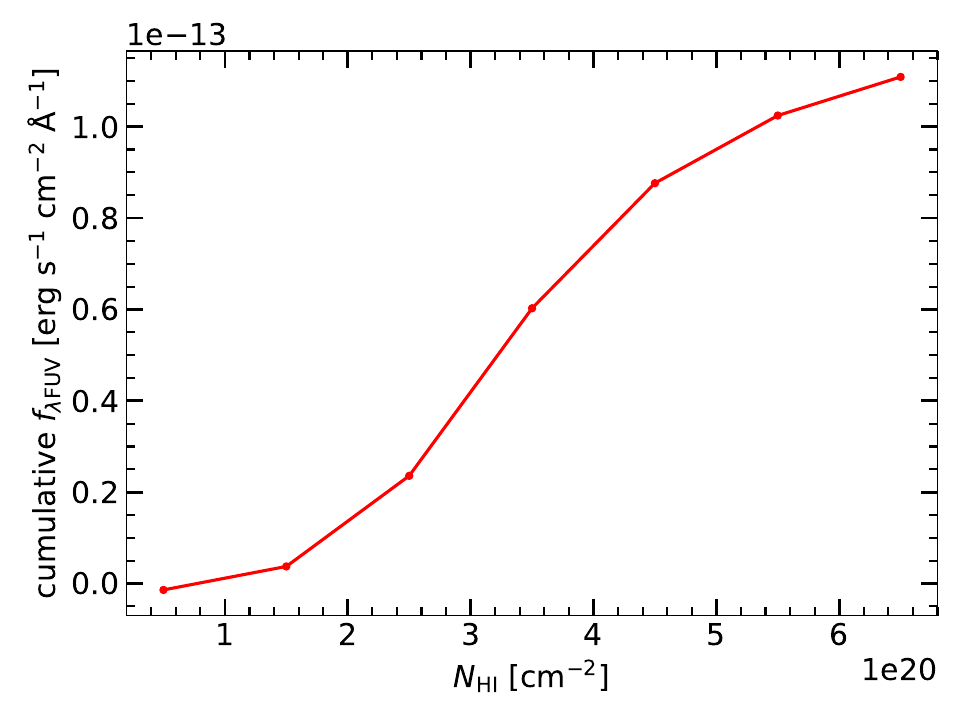}
  \caption{Cumulative FUV flux compared against gas density. The flux was counted by summing the fluxes of clusters within the gas density bins and subtracting from this the flux of the OUT clusters scaled to the size of the bin.}
  \label{fig:gasbins}
\end{figure}

To test if all the clusters associated with M83 are within our IN area, we separated the outer H{~\sc i} disk into seven regions according to $N_{\textrm{H{~\sc i}}}$ (each region covered a range of $10^{20}$ cm$^{-2}$ in $N_{\textrm{H{~\sc i}}}$) and calculated the total FUV flux of objects within each region after subtracting the FUV flux of the OUT objects scaled to the spatial size of the region. This is shown in Fig. \ref{fig:gasbins}. We found that after subtracting the scaled background-source flux, the region with the lowest gas density (0--10$^{20}$ cm$^{-2}$) has $f_{\lambda\textrm{FUV}} \sim 0$, meaning all of the clusters within this region are background sources. This indicates that there are no clusters associated with M83 outside our IN area.

\subsection{Detection limit and completeness}

To determine the FUV detection limit of our catalog, we performed a series of completeness tests for different FUV magnitudes. We did this by selecting 100 background boxes of 20~px $\times$ 20~px in size where Astrodendro found no objects in the background-subtracted FUV image, inserting artificial point sources into these background boxes, and running Astrodendro over the boxes. We used the \emph{GALEX} FUV PSF for the shape of the artificial objects and applied random Poisson noise to each object. We repeated this with different point source magnitudes between 23 mag and 24 mag and ran 10\,000 Poisson noise initializations for each magnitude. For FUV $= 23.5$ mag, we recovered a fraction $f_\textrm{r} = 0.93 \pm 0.03$ of the artificial objects with Astrodendro using the same detection parameters we used for our catalog, while for FUV $=24$ mag, we obtained a recovery fraction of $f_\textrm{r} = 0.66 \pm 0.05$. We adopted FUV $= 23.8$ ($f_\textrm{r} = 0.78 \pm 0.04$), which is equal to the FUV magnitude of a single B0 star at the distance of M83 \citepalias{koda2012imf}, as the detection limit of our catalog.

\subsection{Uncertainty in flux measurements}
\label{sec:error}
There are several sources of uncertainty in our flux measurements, the most significant of which are the Poisson noise, the uncertainty in the extinction value, and the uncertainty in our photometric calibration. We ignored the photometric calibration uncertainty, as the NB\_659 filter throughput varies less than 0.05 mag over the OmegaCAM FoV \citep{drew2014filter}. We estimated the Poisson noise by finding the standard deviation in the background ($\sigma_{\textrm{bg}}$), which when quadratically summed with the source Poisson noise gives the error in flux as 

\begin{equation}
\sigma_{\textrm{Poisson}}^2 = \frac{F}{g_{\textrm{eff}}} + \sigma_{\textrm{bg}}^2 N_{\textrm{px}},
\end{equation}

\noindent
where $F$ is the FUV or H$\alpha$ flux, $N_{\textrm{px}}$ is the number of pixels for the object in question, and $g_\textrm{eff}$ is the effective gain. The effective gain was obtained as $g_\textrm{eff} = g \times t_\textrm{exp} / f$, where $g$ is the instrumental gain, which is one for \emph{GALEX} and 0.5 for OmegaCAM; $t_\textrm{exp}$ is the total exposure time; and $f$ is the conversion factor from counts to flux units.

To obtain the error in flux due to uncertain extinction, we used a Monte Carlo method in which the extinctions reported by \citet{bresolin2009spec} were randomly sampled and an additional extinction correction was applied to each of our objects following the \citet{pei1992dust} extinction curve for the SMC. We chose to use the extinctions of \citet{bresolin2009spec} as the basis of this estimate rather than those of \citet{gildepaz2007spec} since \citet{bresolin2009spec} have a larger sample and deeper spectra. We rejected the largest extinction reported by \citet{bresolin2009spec} ($A_V = 1.16$) as an outlier. We then took the one-sided 68th percentiles around the median of the flux distribution given by the Monte Carlo method as the errors. Since varying the assumed extinction only affects the extinction-corrected fluxes toward the positive direction, the resulting errors are very asymmetric, and the mean values of the Monte Carlo flux distribution are heavily biased. We did not take this bias into account when reporting fluxes and derived values, as we instead examine the effects of higher internal extinction in Sect. \ref{sec:dext}. All the fluxes and quantities derived from fluxes we report have been corrected for $A_V^\textrm{MW}$ and $A_V^\textrm{int}$ as described in Sect. \ref{sec:ext}, while all flux errors and errors of quantities derived from the fluxes contain the $\sigma_{\textrm{Poisson}}$ and the errors obtained from the Monte Carlo method.

\subsection{Luminosity functions}
\begin{figure}
  \centering
  \includegraphics[width=0.5\textwidth]{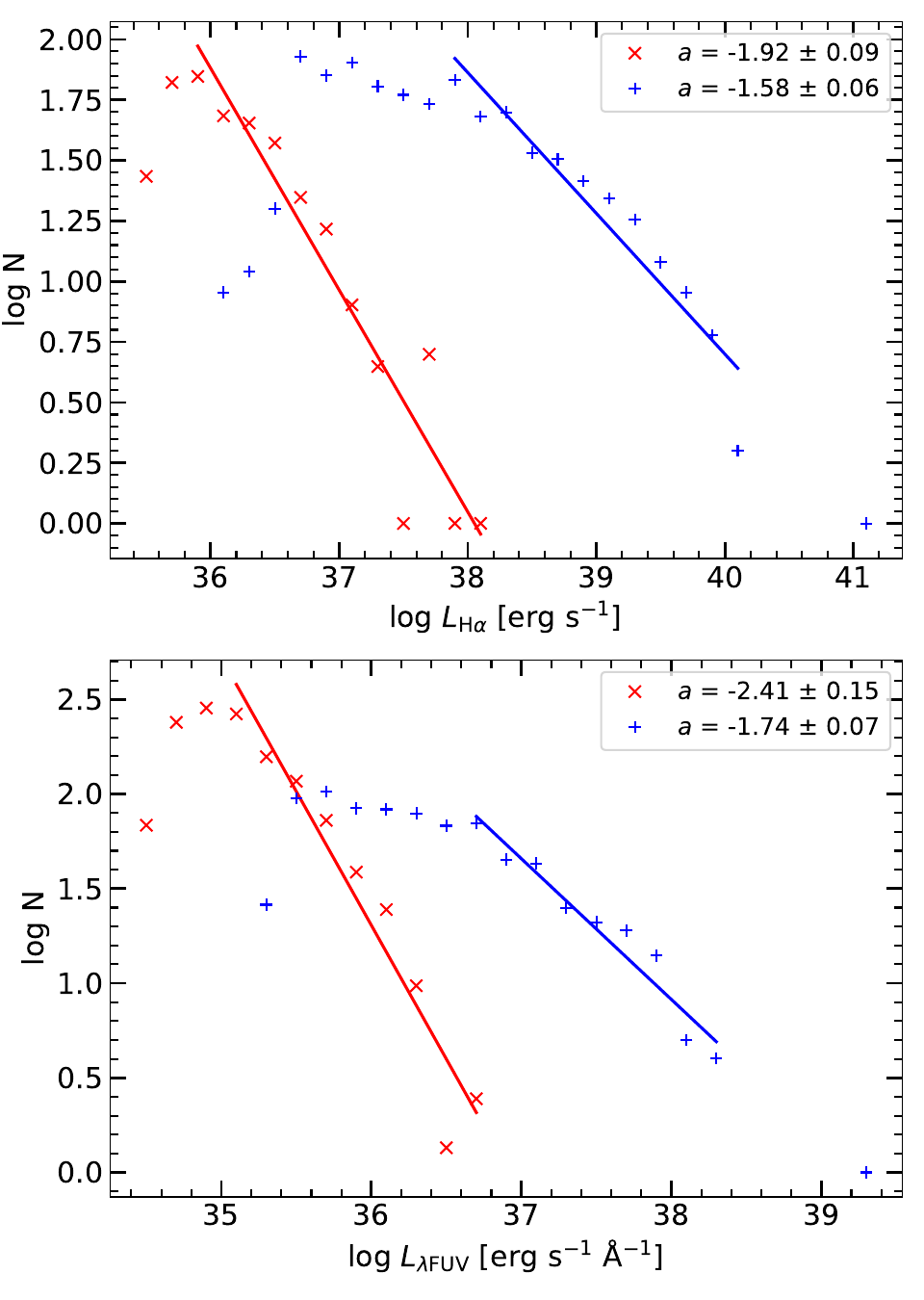}
  \caption{Luminosity functions for M83. The red crosses correspond to outer disk objects and the blue plus signs to inner disk objects. The solid lines indicate the best fit and the range in log $L$ over which the fit was made. The slopes of the best fits ($a$) are given in the legend. \emph{Top panel:} H$\alpha$ luminosity functions. \emph{Bottom panel:} FUV luminosity functions.}
  \label{fig:lf}
\end{figure}

To investigate if the distributions of star clusters and H{~\sc ii} regions in the M83 outer disk differ from typical cluster and H{~\sc ii} region populations, we constructed H$\alpha$ and FUV luminosity functions (LFs) of our sample objects. In order to find any differences between the outer and inner disk, we also constructed H$\alpha$ and FUV LFs for the inner disk clusters in M83. To do this, we created a catalog of inner disk objects by running Astrodendro over the \emph{GALEX} FUV image and again selecting the leaves of the dendrogram as the star clusters using the same minimum pixel value, minimum significance, and minimum size as for the outer disk catalog. We obtained the H$\alpha$ fluxes of the inner disk clusters again with forced photometry. As extinction is more significant in the dusty inner parts of galaxies, we used the total extinction values of $A_{\textrm{FUV}} = 2$ mag and $A_{\textrm{H}\alpha} = 1.4$ mag reported by \citet{boissier2005ext} for the inner disk extinction correction in M83. The LFs are shown in Fig. \ref{fig:lf}.

For the H$\alpha$ LF in the outer disk, we divided the objects with a 5$\sigma$ H$\alpha$ detection in our sample into bins of 0.2 in the log of the luminosity and fit a function of type $N(L_{\textrm{H}\alpha}) = AL_{\textrm{H}\alpha}^a$ to the high luminosity side of it. The shallowing of the LF in the low luminosity side is caused by observational effects, such as missing clusters near the sensitivity limit and the blending of objects, and due to this, we did not include these bins in the fit. We subtracted the number of OUT objects scaled with the ratio of areas from the IN objects for each bin to account for the background contamination. For the slope of the LF above log $L_{\textrm{H}\alpha}=35.8$ erg s$^{-1}$, we obtained $a=-1.92 \pm 0.09$. Using the same bin size and the same type of function, we found a slope of $a=-1.58 \pm 0.06$ for the H$\alpha$ LF above log $L_{\textrm{H}\alpha}=37.8$ erg s$^{-1}$ in the inner disk. In addition to the low luminosity side of the LF, for the inner disk, we also excluded the highest luminosity bin from the fit, which corresponds to the central starburst region of M83. The given errors are 1$\sigma$ fitting errors. Both of these values fall within the range of typical values for an H{~\sc ii} region population \citep{kennicutt1989lf}. The slope for the outer disk H{~\sc ii} regions is steeper than for the inner disk H{~\sc ii} regions, and the regions in the inner disk are brighter overall. A similar, although much subtler, trend exists between interarm and arm H{~\sc ii} regions, with luminous H{~\sc ii} regions being rarer between arms than within arms \citep{knapen1993lf1,knapen1998lf2}.

We used the same bin size and same type of function when constructing the FUV LFs, but we did not restrict the LFs to only objects with an H$\alpha$ detection. After subtracting the scaled number of OUT objects from the IN objects in each bin, we obtained $a=-2.41 \pm 0.15$ for the FUV LF slope above log $L_{\lambda\textrm{FUV}}=35.0$ erg s$^{-1}$ Å$^{-1}$ in the outer disk. For the inner disk FUV LF slope above log $L_{\lambda\textrm{FUV}}=36.6$ erg s$^{-1}$ Å$^{-1}$, we found $a=-1.74 \pm 0.07$. Similar to the H$\alpha$ LFs, the slope of the LF is steeper for the outer disk clusters, and overall they are less luminous than the inner disk clusters. A difference in slope between the inner disk FUV LF and outer disk FUV LF was already reported by \citet{thilker2005m83}.

We note that the blending of objects is considerable in our inner disk catalog, as indicated by the apparent lack of low luminosity objects therein. To test this blending, we also built an LF of the inner disk H{~\sc ii} regions using the unconvolved H$\alpha$ data with the original $1\farcs0$ resolution. This gave a steeper $a=-1.84 \pm 0.05$ slope, confirming that the blending causes some shallowing of the LF slope. However, even with the higher resolution, the slope is still shallower than the outer disk slope, lending credence to the idea that the difference in the slope between the inner and outer disk objects may be physical. The difference in cluster brightness between the outer and inner disk may also be affected by uncertainties in the extinction correction. However, the $\sim 2.5$ mag difference seen in Fig. \ref{fig:lf} cannot be completely explained by the uncertainties in the extinction correction, as it is much greater than the $< 1.5$ mag differences in our outer and inner disk extinction corrections. This indicates that there are fewer massive star clusters and H{~\sc ii} regions in the outer disk compared to the inner disk.

\subsection{H$\alpha$-to-FUV coincidence}
As H$\alpha$ maps the emission of ionized gas and FUV maps starlight, they may not always spatially coincide. This becomes more likely as stellar clusters age and the most massive stars explode as supernovae, potentially disrupting the ionized gas envelope around the cluster \citep{churchwell2006bubble, whitmore2011hii}. To investigate the coincidence of H$\alpha$ and FUV emission in our sample, we computed the pixel-to-pixel Spearman rank correlation between H$\alpha$ flux and FUV flux for all of our objects with a 5$\sigma$ H$\alpha$ detection. We selected as correlated objects where the correlation is significant to a 0.01 level. As the H$\alpha$-to-FUV flux ratio ($F_{\textrm{H}\alpha}/f_{\lambda \textrm{FUV}}$) correlates with the age of the stellar population, if older clusters are more likely to have disrupted ionized gas envelopes and non-coincidental H$\alpha$ and FUV emission, we should find higher $F_{\textrm{H}\alpha}/f_{\lambda \textrm{FUV}}$ in correlated objects. Summing the fluxes over the correlated objects and subtracting the scaled flux of OUT objects from the flux of IN objects, we found a total log$(F_{\textrm{H}\alpha}/f_{\lambda\textrm{FUV}}) = 0.95 \pm 0.04$ in units of log(Å), whereas when summing over the non-correlated objects and subtracting the background, we found a total log$(F_{\textrm{H}\alpha}/f_{\lambda\textrm{FUV}}) = 0.58 \pm 0.03$, where the errors were obtained with the Monte Carlo method. This indicates that the clusters where H$\alpha$ and FUV correlate are indeed younger than the clusters where they do not.

\section{Stellar population synthesis modeling}
\label{sec:models}

\begin{figure*}
  \centering
  \includegraphics[width=\textwidth]{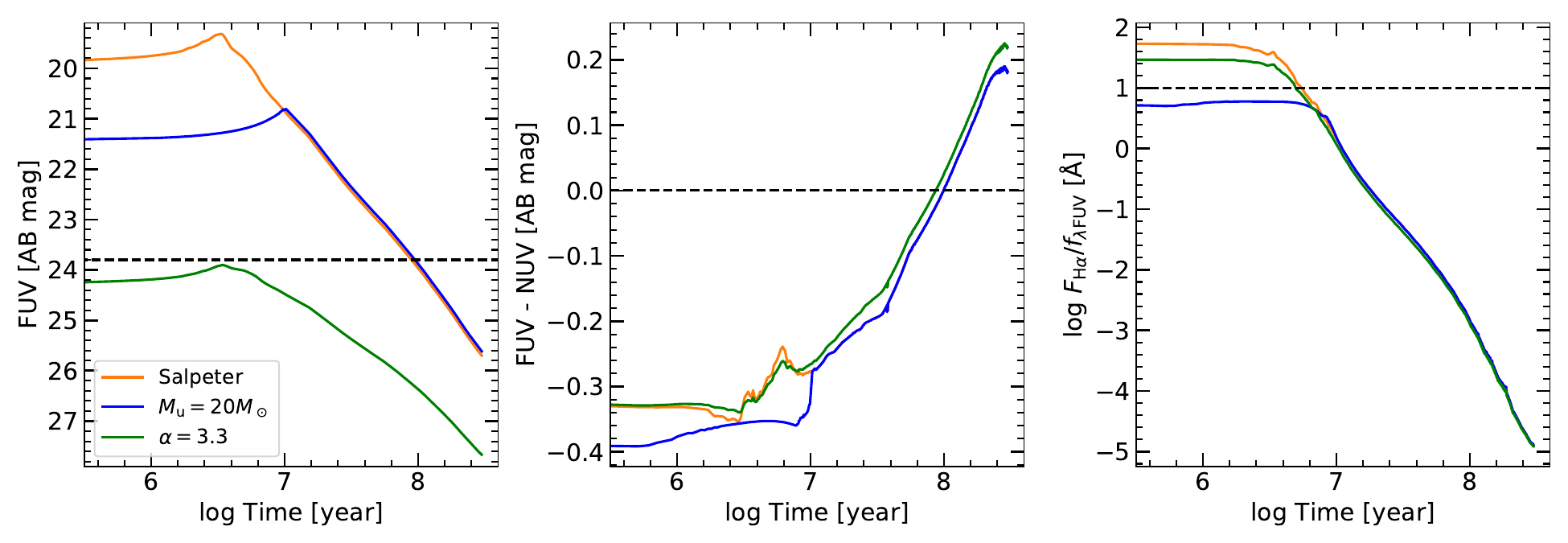}
  \caption{Photometric evolution of our instantaneous burst \textsc{Starburst99} models. The standard \citet{salpeter1955imf} IMF is shown in orange, while a truncated IMF (at $20 M_{\odot}$) is shown in blue, and a steep IMF ($\alpha = 3.3$) is shown in green. \emph{Left panel:} Time evolution of FUV magnitude for a cluster mass of $2000 M_{\odot}$. The horizontal dashed line shows the detection limit for our sample (FUV = 23.8 mag). \emph{Middle panel:} Time evolution of FUV$-$NUV color. The horizontal dashed line shows the cut for blue clusters (FUV$-$NUV = 0). \emph{Right panel:} Time evolution of log($F_{\textrm{H}\alpha} / f_{\lambda \textrm{FUV}}$). The horizontal dashed line shows the cut for H$\alpha$-bright clusters (log$(F_{\textrm{H}\alpha}/f_{\lambda \textrm{FUV}}) = 1$).}
  \label{fig:models}
\end{figure*}

\subsection{Starburst99 models}
To gain insight into the physical processes involved in the formation and evolution of the stellar populations in the outer disk, we constructed stellar population synthesis models using the code \textsc{Starburst99} \citep{leitherer1999sb99}. We adopted the Padova stellar evolution tracks with asymptotic giant branch stars and low metallicity ($0.2 Z_{\odot}$).

\begin{table}
  \caption{Model parameters}
  \label{tab:models}
  \centering
  \begin{tabular}{c c c c c}
    \hline\hline
    Model set & \multicolumn{2}{c}{IMF} & Tracks & $Z/Z_{\odot}$ \\
     & $\alpha$ & $M_{\textrm{u}}/M_\odot$ & & \\
    \hline
    Salpeter & 2.35 & 100 & Padova & 0.2 \\
    Truncated & 2.35 & 20--90 & Padova & 0.2 \\
    Steep & 2.5--3.3 & 100 & Padova & 0.2 \\
    \hline
  \end{tabular}
  \tablefoot{The \textsc{Starburst99} Padova stellar evolution tracks including asymptotic giant branch stars were used.}
\end{table}

We compared the standard Salpeter IMF \citep{salpeter1955imf} to truncated models and to models with a steeper IMF slope. The form of the IMF we used is

\begin{equation}
\xi(m) = \xi_0 m^{-\alpha} \textrm{~~~~~for~~~~~} 0.1 \textrm{~M}_{\odot} < m < M_\textrm{u},
\end{equation}

\noindent
where $m$ is the stellar mass in units of solar masses. For the Salpeter IMF, the power-law index ($\alpha$) is 2.35, and the upper mass limit ($M_{\textrm{u}}$) is $100 M_{\odot}$. For the models with alternative IMFs, we varied $M_{\textrm{u}}$ and $\alpha$ separately, constructing one set of models with $M_{\textrm{u}}$ varied in $10 M_\odot$ steps from $M_{\textrm{u}}=20 M_\odot$ to $M_{\textrm{u}}=90 M_\odot$ and another set of models with $\alpha$ varied in 0.1 steps from $\alpha=2.5$ to $\alpha=3.3$. The parameters of our models are gathered in Table \ref{tab:models}.

We constructed models with two kinds of SFHs, namely, instantaneous burst or continuous star formation, to study the temporal evolution of the stellar populations in the outer disk. An instantaneous burst is a good approximation for a single stellar cluster, while if a constant cluster formation rate is assumed, a model with continuous star formation better represents the integrated emission of the entire outer disk. We used a cluster mass of $2000 M_{\odot}$ for our instantaneous burst models and an SFR of $1 M_{\odot}$ yr$^{-1}$ for our continuous star formation models. We note that the cluster mass and SFR have no impact on the colors and flux ratios of our models, as long as the cluster mass is high enough ($m_{\textrm{cluster}} \gg 100 M_{\odot}$) to generate the most massive O stars.

\subsection{Modeling cluster counts and completeness limits}
As star clusters age, they become fainter, and eventually they fall below the detection limit of our FUV data ($\sim$23.8 mag). The age when this happens depends on the cluster mass and the IMF. Star clusters also change in color as they age, experiencing a drastic reduction in their H$\alpha$ emission within $\sim 10$ Myr as O stars die, and a more gradual, but nonetheless significant, reduction in their FUV emission within $\sim 100$ Myr as B stars die.

We performed a similar analysis as \citetalias{koda2012imf} by counting H$\alpha$-bright (including O and B stars) and blue (including B stars) clusters in our catalog and comparing them to evolutionary models in order to constrain the IMF at its high-mass end. This required us to use samples that are complete in cluster age, meaning all clusters larger than a limiting mass are observable up to a given age. In other words, in order to compare our models to our data, we had to make sure our sample of star clusters was not biased toward young H$\alpha$-bright clusters due to low-mass clusters remaining above the detection limit only for as long as their O stars live and they are H$\alpha$ bright. This restriction to only massive clusters also ensured that the high-mass end of the IMF would not be stochastically sampled within the sample clusters, thus ruling out stochasticity as the cause of any apparent deviation from the standard Salpeter IMF.

Compared to \citetalias{koda2012imf}, who only examined a model with the Salpeter IMF, we expanded the analysis to several models with different IMFs. As each model with a different IMF also has a different evolutionary sequence, we obtained different completeness limits for each model.

The photometric evolution of our instantaneous burst Salpeter model and the two most extreme instantaneous burst models ($\alpha=2.35$, $M_{\textrm{u}} = 20 M_\odot$; and $\alpha = 3.3$, $M_{\textrm{u}} = 100 M_\odot$) are shown in Fig. \ref{fig:models}. In the leftmost panel, the FUV flux corresponds to a cluster mass of $2\times 10^3 M_\odot$, while the other two panels show flux ratios and are thus valid for all cluster masses. We selected our completeness limits to include only clusters that are massive enough to remain visible in our data (FUV < 23.8 mag) for at least 100 Myr. As demonstrated by the middle panel, this age constraint corresponds for all models to the age when clusters appear blue in UV (FUV$-$NUV < 0.0 mag), or equivalently, to the lifetime of B stars. For the model with a Salpeter IMF and the truncated models, this means a limiting cluster mass of $\sim 2 \times 10^3 M_\odot$, as can be seen from the left panel of Fig. \ref{fig:models}, and for the steep models, this means limiting cluster masses between $\sim 2 \times 10^3$ and $\sim2 \times 10^4 M_\odot$. We defined H$\alpha$-bright clusters as those with log$(F_{\textrm{H}\alpha}/f_{\lambda \textrm{FUV}}) > 1$. Clusters with H$\alpha$-to-FUV flux ratio below this do not have O stars. For example, this applies to the truncated model with $M_\textrm{u} = 20 M_\odot$ in Fig. \ref{fig:models}.

Using the reasonable assumptions of instantaneous cluster formation and a constant cluster formation rate, the relative number of H$\alpha$-bright clusters to blue clusters ($N_{\textrm{H}\alpha} / N_{\textrm{blue}}$) can be obtained from the times that the clusters remain H$\alpha$ bright and blue ($t_{\textrm{H}\alpha}$ and $t_{\textrm{blue}}$). Under these assumptions, the number of clusters younger than a certain age should be proportional to that age. Thus, we expected $N_{\textrm{H}\alpha}/N_{\textrm{blue}} = t_{\textrm{H}\alpha}/t_{\textrm{blue}}$.

Based on Fig. \ref{fig:models}, we found that the standard Salpeter IMF and the truncated IMFs create clusters with $t_{\textrm{blue}} = 99$ Myr, while clusters with steep IMFs remain blue for a little less time, down to $t_{\textrm{blue}} =  87$ Myr with the $\alpha = 3.3$ IMF. Clusters with steep IMFs also remain H$\alpha$ bright for slightly shorter times than clusters with a Salpeter IMF, down to $t_{\textrm{H}\alpha} = 5.0$ Myr for $\alpha = 3.3$ compared to $t_{\textrm{H}\alpha} = 5.5$ Myr for the Salpeter IMF. Thus, $t_{\textrm{H}\alpha}/t_{\textrm{blue}} = 0.056$ for the Salpeter IMF, and $t_{\textrm{H}\alpha}/t_{\textrm{blue}} = 0.058$ for the $\alpha = 3.3$ IMF. Clusters with the IMF truncated at $M_{\textrm{u}} = 20 M_\odot$, on the other hand, are never H$\alpha$ bright. As H$\alpha$-bright clusters exist in the M83 outer disk, strongly truncated models do not represent observations well. For all the models that produce H$\alpha$-bright clusters, the resulting expected $N_{\textrm{H}\alpha}/N_{\textrm{blue}} = 0.06$ is nearly invariant with respect to the IMF.

\section{Results}
\label{sec:results}

\subsection{Observed cluster counts in the M83 outer disk}
\begin{figure}
  \centering
  \includegraphics[width=0.5\textwidth]{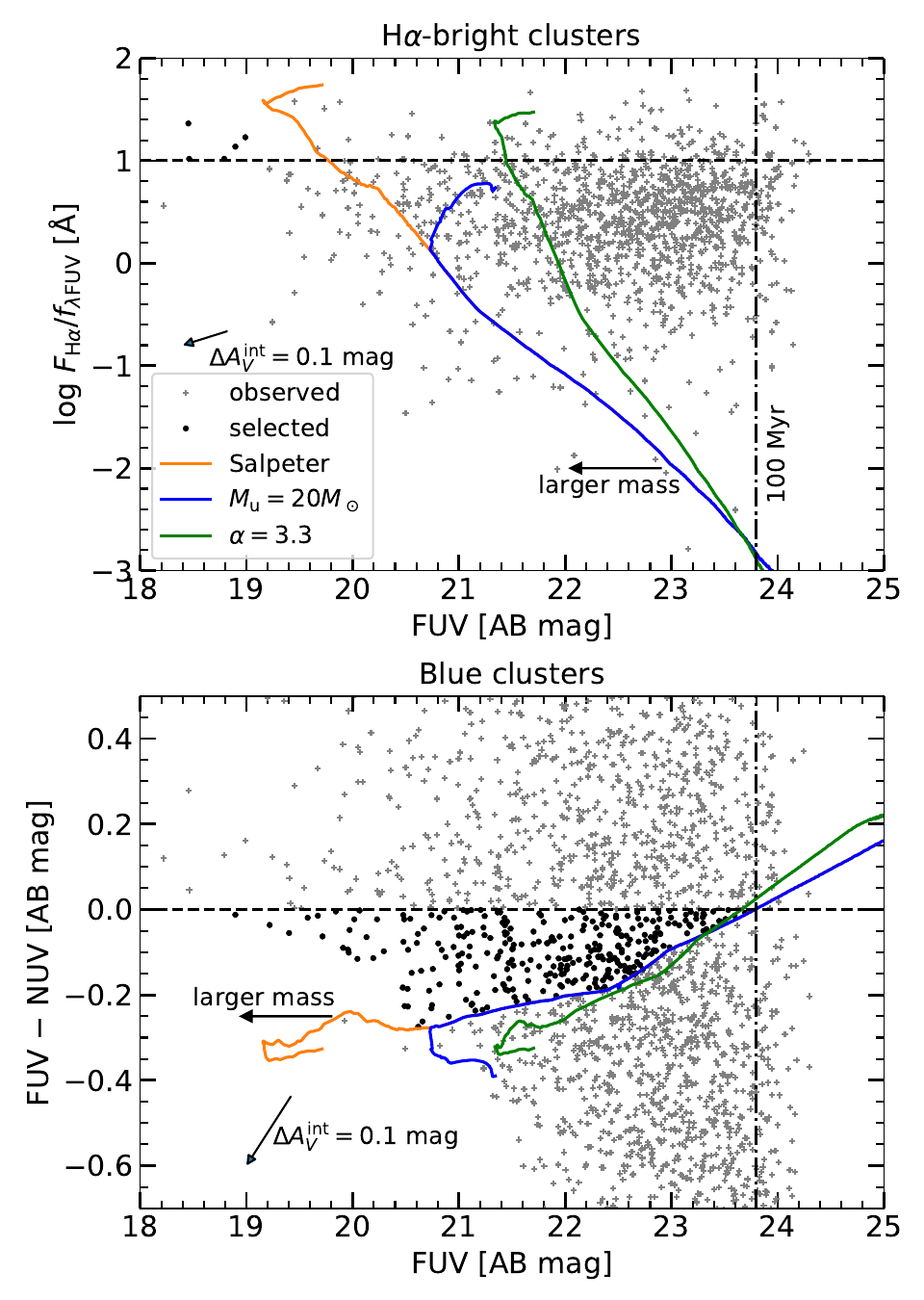}
  \caption{Color-magnitude diagrams for the IN objects. \emph{Top panel:} Comparison of log($F_{\textrm{H}\alpha} / f_{\lambda\textrm{FUV}}$) vs. FUV magnitude. The model evolutionary sequences for the standard Salpeter IMF (orange) and the most extreme truncated ($M_{\textrm{u}} = 20 M_\odot$; blue) and steep ($\alpha = 3.3$; green) IMFs are shown as solid curves. The mass of each model cluster has been normalized so that they reach an age of 100 Myr at a FUV = 23.8 mag (marked with a vertical dot-dashed line). The direction of increasing cluster mass for the models is shown with a horizontal black arrow. The horizontal dashed line indicates the demarcation for H$\alpha$-bright clusters. The H$\alpha$-bright clusters that are also above the Salpeter mass completeness limit are marked with black dots. The observed points have been corrected for $A_V^{\textrm{int}}$ measured from the H{~\sc i} data and $A_V^{\textrm{MW}}$. The change in the position of observed points if $A_V^{\textrm{int}}$ is assumed to be larger by 0.1 mag is also shown with a black arrow. \emph{Bottom panel:} Same as the top panel but for FUV$-$NUV vs. FUV and with blue clusters marked with black dots. We used CMDs to find the H$\alpha$-bright and blue clusters in our complete samples.}
  \label{fig:cmd}
\end{figure}

To obtain the number of H$\alpha$-bright and blue clusters in the M83 outer disk, we selected clusters in color-magnitude space with masses above our completeness limits and log$(F_{\textrm{H}\alpha}/f_{\lambda \textrm{FUV}}) > 1$ or FUV$-$NUV < 0.0, respectively. Color-magnitude diagrams (CMDs) for IN objects are shown in Fig. \ref{fig:cmd}. Overlaid as colored curves are the model evolutionary sequences for the standard Salpeter IMF and the most extreme truncated and steep IMFs ($M_{\textrm{u}} = 20 M_\odot$ and $\alpha=3.3$). The model cluster masses are normalized so that they reach 100 Myr age at FUV = 23.8 mag ($m_\textrm{cluster} \approx 2 \times 10^3 M_\odot$ for the Salpeter and the truncated model and  $m_\textrm{cluster} \approx 2 \times 10^4 M_\odot$ for the steep model). For each model, we selected clusters that are left of the overlaid evolutionary sequence curves and above the limiting log$(F_{\textrm{H}\alpha}/f_{\lambda \textrm{FUV}}) > 1$ or below the limiting FUV$-$NUV < 0.0 mag to obtain our cluster counts ($N_{\textrm{H}\alpha}$, $N_{\textrm{blue}}$, respectively). The upper panel of Fig. \ref{fig:cmd} illustrates well how the different IMFs affect the completeness limits, as we observed four times the amount of clusters (gray and black dots in Fig. \ref{fig:cmd}) above the limiting mass for the $\alpha = 3.3$ IMF (left of the green curve) compared to the amount of clusters above the limiting mass for the Salpeter IMF (left of the red curve).

\begin{figure}
  \centering
  \includegraphics[width=0.5\textwidth]{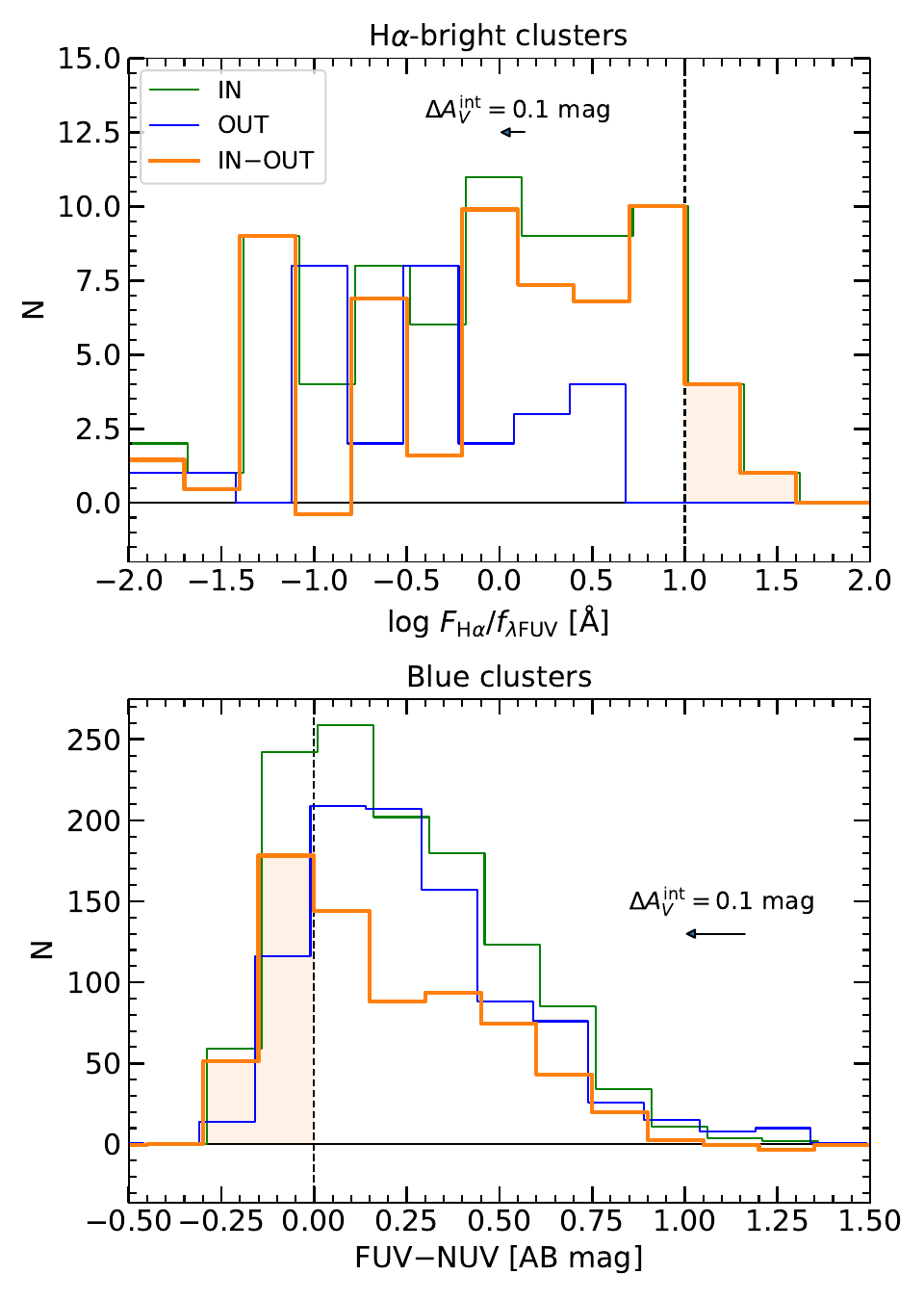}
  \caption{Histograms of the extinction-corrected flux ratios and UV colors of mass-selected objects assuming a Salpeter IMF. \emph{Top panel:} Histogram of log($F_{\textrm{H}\alpha}/f_{\lambda\textrm{FUV}}$). The green histogram shows the objects within the H{~\sc i} gas footprint, the blue histogram shows the objects outside it, and the orange histogram shows the star clusters associated with M83. The orange histogram was obtained by scaling the blue histogram with the IN/OUT area ratio and subtracting it from the green histogram. The change in the flux ratio if $A_V^{\textrm{int}}$ is assumed to be larger by 0.1 mag is shown with a black arrow. \emph{Bottom panel:} Same as the top panel but for FUV$-$NUV. The shaded parts of the histograms correspond to the H$\alpha$-bright (log($F_{\textrm{H}\alpha}/f_{\lambda\textrm{FUV}}$) > 1) clusters and the blue (FUV$-$NUV < 0) clusters, respectively. These histograms illustrate the subtraction of background contamination from our cluster counts.}
  \label{fig:hist}
\end{figure}

To remove the background contamination, we selected clusters in both our IN sample and OUT sample and subtracted the number of H$\alpha$-bright clusters and blue clusters among OUT objects, scaled with the IN/OUT area ratio, from those among IN objects. This process is demonstrated in histograms in Fig. \ref{fig:hist}.

\begin{figure*}[ht]
  \centering
  \includegraphics[width=\textwidth]{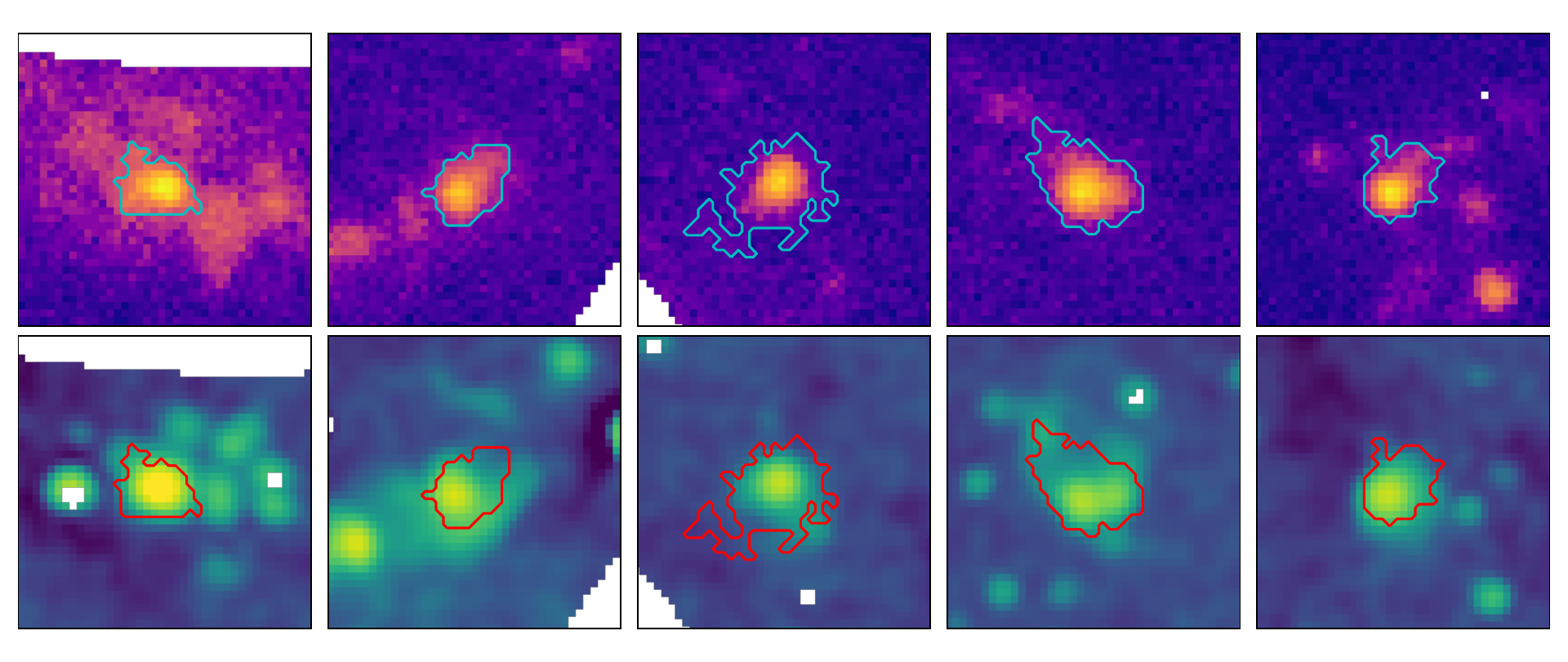}
  \caption{Postage stamp images of the five most massive H$\alpha$-bright star clusters in the M83 outer disk. \emph{Top row:} \emph{GALEX} FUV images. \emph{Bottom row:} OmegaCAM H$\alpha$ images. The isophotal aperture used for photometry is contoured in each image. These clusters correspond to the H$\alpha$-bright clusters in the mass-selected sample when assuming a Salpeter IMF. Each of the images is $1\arcmin \times 1\arcmin$ in size.}
  \label{fig:stamps}
\end{figure*}

\begin{figure}
  \centering
  \includegraphics[width=0.5\textwidth]{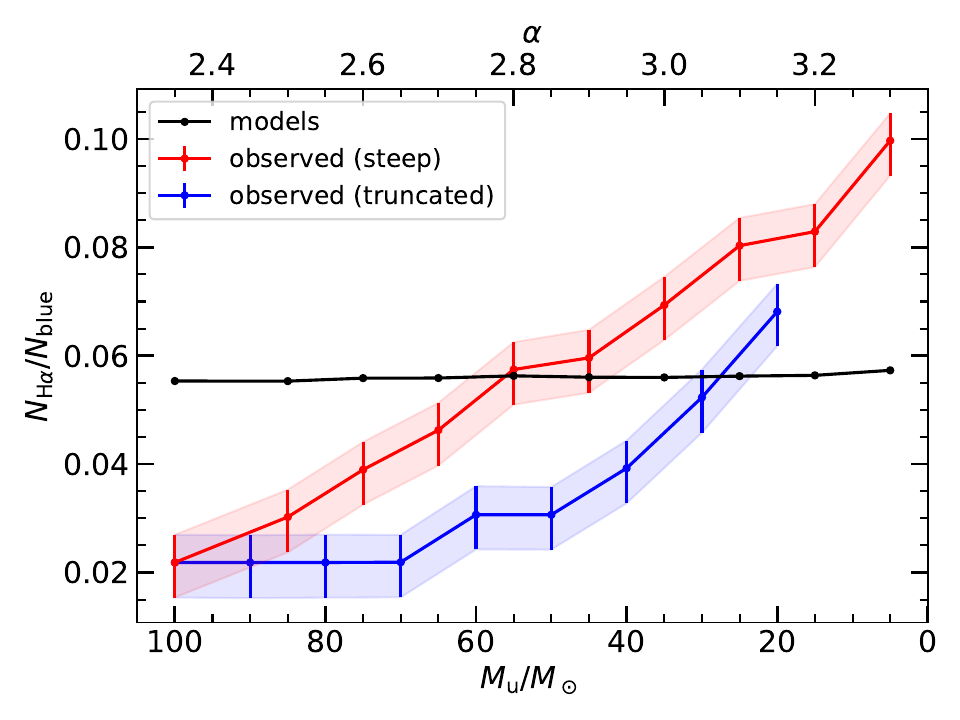}
  \caption{Observed and modeled $N_{\textrm{H}\alpha}/N_{\textrm{blue}}$ for different IMFs. The black curve show values predicted by our steep models with different IMF slopes ($\alpha$). These were obtained as $N_{\textrm{H}\alpha}/N_{\textrm{blue}} = t_{\textrm{H}\alpha}/t_{\textrm{blue}}$ while assuming instantaneous cluster formation and a constant cluster formation rate. The values predicted by the truncated models are not shown for clarity, but they are very close to the values predicted by the steep models. The first dot from the left is given by the model with the standard Salpeter IMF ($\alpha = 2.35$). The red curve indicates the observed value in the M83 outer disk using different mass selection criteria imposed by IMFs with different $\alpha$. The blue curve is the same as the red curve but for the truncated IMFs with different $M_{\textrm{u}}$. The shaded areas around the lines show the errors in the observed values. The observations and models agree for $2.8 \le \alpha \le 2.9$, or $M_{\textrm{u}} \approx 30 M_\odot$.}
  \label{fig:avr}
\end{figure}

In our mass-selected sample assuming a Salpeter IMF, counting from Fig. \ref{fig:hist} (orange), we found $N_{\textrm{H}\alpha} = 5 \pm 2$ and $N_{\textrm{blue}} = 229^{+12}_{-13}$, giving $N_{\textrm{H}\alpha} / N_{\textrm{blue}} = 0.02 \pm 0.01$. The five most massive H$\alpha$-bright clusters, corresponding to all H$\alpha$-bright clusters of the mass-selected sample assuming a Salpeter IMF, are shown in Fig. \ref{fig:stamps}. We counted the blue and H$\alpha$-bright clusters in a similar manner using completeness limits based on the steep and truncated IMF models. Figure \ref{fig:avr} shows how the completeness limits assuming different IMFs affect $N_{\textrm{H}\alpha} / N_{\textrm{blue}}$.  The observed cluster counts assuming an IMF slope $2.8 \le \alpha \le 2.9$ agree best with the $N_{\textrm{H}\alpha} / N_{\textrm{blue}} \approx 0.06$  predicted by models, while observed cluster counts assuming IMFs with a steeper or shallower slope give too high or too low $N_{\textrm{H}\alpha} / N_{\textrm{blue}}$, respectively. There is a similar, albeit shallower, trend among truncated IMFs where lower than predicted $N_{\textrm{H}\alpha} / N_{\textrm{blue}}$ is found when assuming $M_{\textrm{u}} > 30 M_\odot$ and higher than predicted $N_{\textrm{H}\alpha} / N_{\textrm{blue}}$ is found when assuming $M_{\textrm{u}} \approx 20 M_\odot$. Nonetheless, the best agreement between the observed cluster counts and the truncated model predictions is found for $M_{\textrm{u}} \approx 30 M_\odot$.

\subsection{H$\alpha$-to-FUV flux ratio}
\begin{figure}
  \centering
  \includegraphics[width=0.5\textwidth]{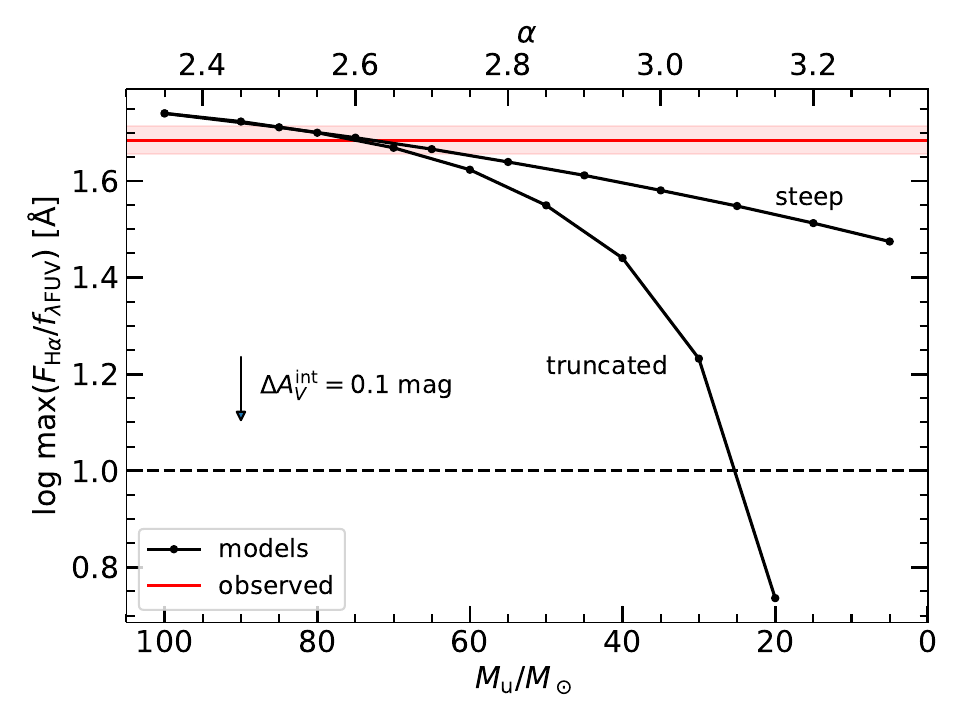}
  \caption{Maximum H$\alpha$-to-FUV flux ratios of stellar clusters predicted by our models. The black curves show $F_{\textrm{H}\alpha}/f_{\lambda \textrm{FUV}}$ against IMF slope ($\alpha$) and truncation mass ($M_{\textrm{u}}$) for the set of steep models and the set of truncated models, respectively. The first dot from the left is from the model with the standard Salpeter IMF ($\alpha = 2.35$, $M_\textrm{u} = 100 M_\odot$). The horizontal red line and shaded area show the maximum observed $F_{\textrm{H}\alpha}/f_{\lambda \textrm{FUV}}$ in our catalog and its errors. The dashed black horizontal line is the limit of H$\alpha$ brightness that we adopted. The change in the observed flux ratio if $A_V^{\textrm{int}}$ is assumed to be larger by 0.1 mag is shown with a black arrow. Only models with $\alpha < 2.8$ or $M_{\textrm{u}} > 60 M_\odot$ can generate clusters with as high a $F_{\textrm{H}\alpha}/f_{\lambda \textrm{FUV}}$ as are observed.}
  \label{fig:maxfr}
\end{figure}

Another way to constrain the high-mass end of the IMF is to look at the H$\alpha$-to-FUV flux ratio. As already shown in Fig. \ref{fig:models}, the $M_{\textrm{u}} = 20 M_\odot$ IMF is incapable of producing H$\alpha$-bright clusters. Additionally, no model with an IMF truncated at $M_{\textrm{u}} < 70 M_\odot$ can produce clusters with as high $F_{\textrm{H}\alpha}/f_{\lambda \textrm{FUV}}$ as we observed in the M83 outer disk. Although steep models with $\alpha  > 2.7$ cannot produce clusters as H$\alpha$ bright as the most extreme cases that we observed either, the maximum $F_{\textrm{H}\alpha}/f_{\lambda \textrm{FUV}}$ predicted by models with steep IMFs is still much closer to the observed values than the maximum $F_{\textrm{H}\alpha}/f_{\lambda \textrm{FUV}}$ predicted by models with a low truncation mass. These limitations are illustrated in Fig. \ref{fig:maxfr}.

\begin{figure}
  \centering
  \includegraphics[width=0.5\textwidth]{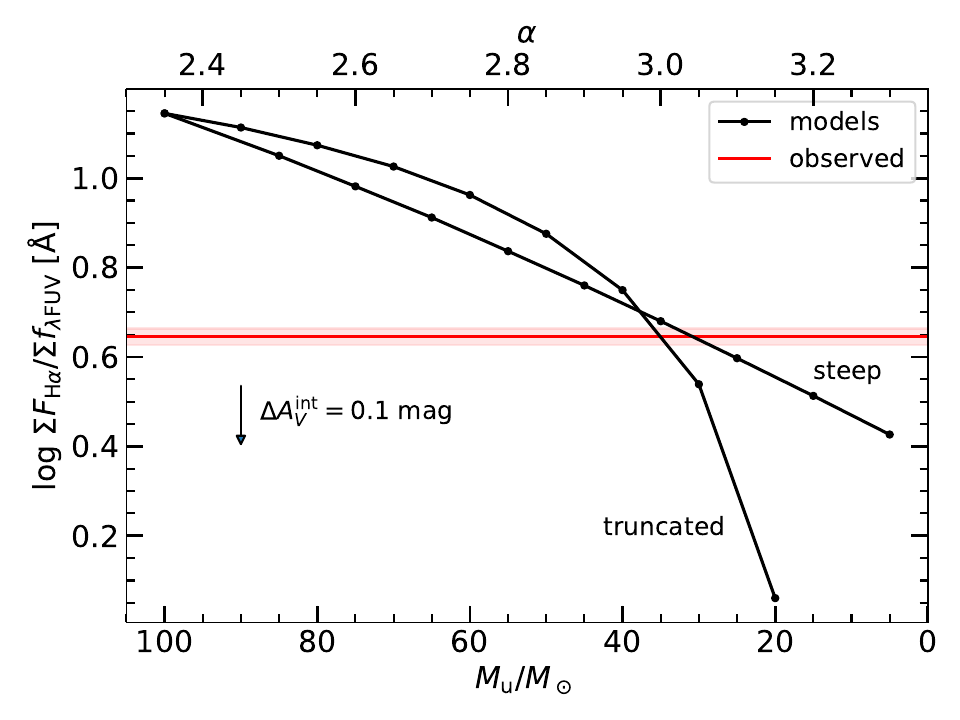}
  \caption{H$\alpha$-to-FUV total flux ratios predicted by our models. The black curves shows values predicted by models with different IMF slopes ($\alpha$) and truncation masses ($M_{\textrm{u}}$) for the set of steep models and the set of truncated models, respectively. The first dot from the left is from the model with the standard Salpeter IMF ($\alpha = 2.35$, $M_\textrm{u} = 100 M_\odot$). The horizontal red line indicates the observed value in the M83 outer disk. The shaded area around the red line shows the error in the observed value. The change in the observed flux ratio if $A_V^{\textrm{int}}$ is assumed to be larger by 0.1 mag is shown with a black arrow. The observations and models agree for $3.0 < \alpha < 3.1$ or $30 M_\odot < M_{\textrm{u}} < 40 M_\odot$.}
  \label{fig:avfr}
\end{figure}

The H$\alpha$-to-FUV total flux ratio over the entire outer disk ($\Sigma F_{\textrm{H}\alpha}/\Sigma f_{\lambda \textrm{FUV}}$) is less sensitive to uncertainties of the colors of individual clusters than $N_{\textrm{H}\alpha} / N_{\textrm{blue}}$ or the maximum $F_{\textrm{H}\alpha}/f_{\lambda \textrm{FUV}}$. Figure \ref{fig:avfr} shows $\Sigma F_{\textrm{H}\alpha}/\Sigma f_{\lambda \textrm{FUV}}$ for our continuous star formation models at 1 Gyr age, well after $\Sigma F_{\textrm{H}\alpha}/\Sigma f_{\lambda \textrm{FUV}}$ reaches an equilibrium if we assume a constant cluster formation rate. We found log$(\Sigma F_{\textrm{H}\alpha}/\Sigma f_{\lambda \textrm{FUV}}) = 1.14$ in units of log(Å) for the standard Salpeter IMF; log$(\Sigma F_{\textrm{H}\alpha}/\Sigma f_{\lambda \textrm{FUV}}) = 0.43$ for the $\alpha = 3.3$ IMF; and log$(\Sigma F_{\textrm{H}\alpha}/\Sigma f_{\lambda \textrm{FUV}}) = 0.06$ for the $M_{\textrm{u}} = 20 M_\odot$ IMF. The flux ratio for our Salpeter IMF model is similar to previous values reported in the literature (e.g., $\sim$1.0 in \citetalias{koda2012imf}; $\sim$1.1 in \citealt{fumagalli2011imf}).

We measured the flux ratio in the M83 outer disk taking into account as much of the flux of the outer disk as possible. To do this, we summed not only the fluxes of our objects but also the fluxes within the associated lowest level dendrogram footprints, and we subtracted from the sums obtained from the IN sample the sums obtained from the OUT sample scaled by the area ratio.\footnote{$\Sigma F_{\textrm{H}\alpha} = 4.9 \times 10^{-13}$ erg s$^{-1}$ cm$^{-1}$ and $\Sigma f_{\lambda \textrm{FUV}} = 1.1 \times 10^{-13}$ erg s$^{-1}$ cm$^{-1}$ Å$^{-1}$ for the catalog objects and additional $\Sigma F_{\textrm{H}\alpha} = 1.1 \times 10^{-13}$ erg s$^{-1}$ cm$^{-1}$ and $\Sigma f_{\lambda \textrm{FUV}} = 2.5 \times 10^{-14}$ erg s$^{-1}$ cm$^{-1}$ Å$^{-1}$ for the lowest level dendrogram footprints.} We obtained log$(\Sigma F_{\textrm{H}\alpha}/\Sigma f_{\lambda \textrm{FUV}}) = 0.64 \pm 0.02$. As can be seen from Fig. \ref{fig:avfr}, this seems to indicate an IMF with a slope between $\alpha=3.0$ and $\alpha=3.1$, or an IMF with a truncation between $M_{\textrm{u}} = 30 M_\odot$ and $M_{\textrm{u}} = 40 M_\odot$, which is apparently in tension with the result we obtained by counting H$\alpha$-bright and blue clusters for the steep models ($2.8 \le \alpha \le 2.9$), and the result we obtained by measuring the maximum $F_{\textrm{H}\alpha} / f_{\lambda\textrm{FUV}}$ for the steep models and the truncated models ($\alpha < 2.8$ or $M_{\textrm{u}} > 60 M_\odot$). However, we show in Sect. \ref{sec:dext} that by assuming an additional internal extinction of $A_V^{\textrm{int}} \approx 0.15$ mag, these results can be reconciled for the steep models.

To confirm a difference between the outer disk and inner disk IMFs, we also measured $\Sigma F_{\textrm{H}\alpha}/\Sigma f_{\lambda \textrm{FUV}}$ over the inner disk. We again used the values of $A_{\textrm{FUV}} = 2$ mag and $A_{\textrm{H}\alpha} = 1.4$ mag reported by \citet{boissier2005ext} for the inner disk extinction in M83. We found log$(\Sigma F_{\textrm{H}\alpha}/\Sigma f_{\lambda \textrm{FUV}}) = 1.26$ for the integrated H$\alpha$ and FUV fluxes within the $5\arcmin$ inner disk, a value much closer to the theoretical Salpeter IMF value of log$(\Sigma F_{\textrm{H}\alpha}/\Sigma f_{\lambda \textrm{FUV}}) = 1.14$, suggesting a considerable difference between the outer disk and inner disk IMFs. We note that our adopted inner disk extinction values suppress FUV more than H$\alpha$ compared to our adopted outer disk extinction values, resulting in a lower extinction-corrected $\Sigma F_{\textrm{H}\alpha}/\Sigma f_{\lambda \textrm{FUV}}$ in the inner disk than it would have been if we would have used our outer disk extinction values everywhere.

We also tested whether $F_{\textrm{H}\alpha}/f_{\lambda \textrm{FUV}}$ in the outer disk was correlated with the radial distance or azimuthal direction related to the center of M83 or to the H{~\sc i} column density. While we found no correlations between $F_{\textrm{H}\alpha}/f_{\lambda \textrm{FUV}}$ and the radial distance or azimuthal direction, we did find a slight trend between $F_{\textrm{H}\alpha}/f_{\lambda \textrm{FUV}}$ and the H{~\sc i} column density. We found this by separating our sample objects into H{~\sc i} column density bins $10^{20}$ cm$^{-2}$ wide  using the LVHIS data and comparing the bin H{~\sc i} column density to the $\Sigma F_{\textrm{H}\alpha}/\Sigma f_{\lambda \textrm{FUV}}$ within the bin. We observed that the H$\alpha$-to-FUV flux ratio increases as the gas density increases, suggesting a potential IMF dependence on gas density. However, aging effects could also be responsible for this trend, as lower density bins may contain proportionally more older clusters that have migrated from higher density bins where star formation density is also higher. To test this, we restricted our sample to the blue clusters that are younger than 100 Myr, and instead of using the H{~\sc i} column density at the position of the cluster, we used the maximum H{~\sc i} column density within a radius that the cluster could have traveled during the last 100 Myr assuming a velocity equal to the H{~\sc i} velocity dispersion in the M83 outer disk ($\sim 15$ km s$^{-1}$; \citealt{heald2016hi}). With this test, the trend vanishes, and we found no correlation between the $F_{\textrm{H}\alpha}/f_{\lambda \textrm{FUV}}$ of the blue clusters and the maximum potential gas density at their formation site.

\subsection{Effect of internal extinction}
\label{sec:dext}
Our cluster count and flux ratio analyses support different types of IMFs for the outer disk of M83: the maximum observed cluster H$\alpha$-to-FUV flux ratio suggests a near-Salpeter IMF, as shown by Fig. \ref{fig:maxfr}, while the number of H$\alpha$-bright and blue clusters together with the H$\alpha$-to-FUV total flux ratio support a steeper or a truncated IMF, as shown by Figs. \ref{fig:avr} and \ref{fig:avfr}. To reconcile these results, we looked into a poorly-constrained parameter in our data, namely, the internal extinction.

\begin{figure}
  \centering
  \includegraphics[width=0.5\textwidth]{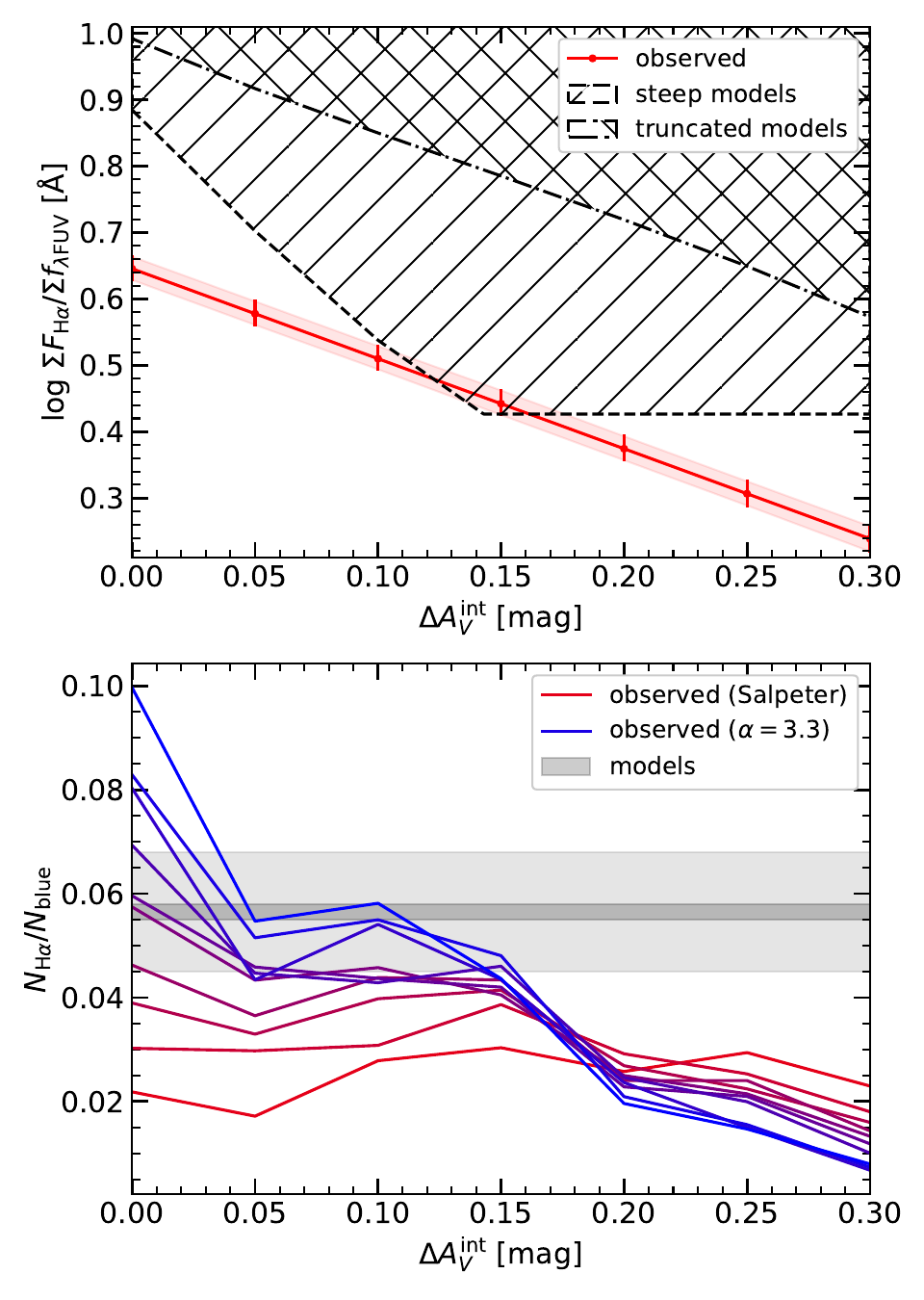}
  \caption{Effect of varying the internal extinction. \emph{Top panel:} Comparison of log($\Sigma F_{\textrm{H}\alpha}/\Sigma f_{\lambda \textrm{FUV}}$) in our catalog vs. the assumed $\Delta A_{V}^{\textrm{int}}$. The dashed curve shows the logarithm of the minimum predicted $\Sigma F_{\textrm{H}\alpha}/\Sigma f_{\lambda \textrm{FUV}}$ among our steep IMF models that are capable of producing clusters with as high a maximum $F_{\textrm{H}\alpha}/f_{\lambda \textrm{FUV}}$ as the maximum $F_{\textrm{H}\alpha}/f_{\lambda \textrm{FUV}}$ observed (southwest to northeast hatches) for each $\Delta A_{V}^{\textrm{int}}$. The dot-dashed curve shows the same for our truncated IMF models (northwest to southeast hatches). The curve for the steep IMF models levels out beyond $\Delta A_{V}^{\textrm{int}} = 0.15$ because at that assumed internal extinction and higher, all of our steep IMF models are capable of producing clusters with a maximum $F_{\textrm{H}\alpha}/f_{\lambda \textrm{FUV}}$ higher than the maximum observed, so the log($\Sigma F_{\textrm{H}\alpha}/\Sigma f_{\lambda \textrm{FUV}}$) value for $\alpha = 3.3$ is given for all $\Delta A_{V}^{\textrm{int}} \ge 0.15$. \emph{Bottom panel:} Comparison of $N_{\textrm{H}\alpha} / N_{\textrm{blue}}$ in our mass-selected catalogs assuming a Salpeter IMF (red curve) and steep IMFs (blue and purple curves) vs. assumed $\Delta A_{V}^{\textrm{int}}$. The values predicted by our steep IMF models are shown by the gray shaded area. The darker shaded area is between the highest and lowest value predicted by  our steep IMF models, and the lighter shaded area is $\pm0.01$ around the darker shaded area, corresponding to the typical errors for $N_{\textrm{H}\alpha} / N_{\textrm{blue}}$ in our analysis. Observations agree with steep IMF models when $0.10 < \Delta A_V^\textrm{int} < 0.15$ mag is assumed.}
  \label{fig:vAv}
\end{figure}

To investigate the effect of the internal extinction on our work, we repeated our analysis using different values of bulk $A_V^\textrm{int}$ on top of the $A_V^\textrm{int}$ calculated from H{~\sc i} data for the extinction correction. This was motivated by the large extinction values measured for individual clusters in \citet{gildepaz2007spec} and \mbox{\citet{bresolin2009spec}}. We varied the additional extinction $\Delta A_V^\textrm{int}$ with 0.05 mag steps from 0.0 mag to 0.3 mag. The effect of $\Delta A_V^\textrm{int}$ on our analysis is demonstrated in Fig. \ref{fig:vAv}. In particular, the upper panel shows the relation of $\Sigma F_{\textrm{H}\alpha}/\Sigma f_{\lambda \textrm{FUV}}$ to the assumed $\Delta A_V^\textrm{int}$. The larger the $\Delta A_V^\textrm{int}$, the smaller the estimated intrinsic H$\alpha$-to-FUV total flux ratio is. As changing the $\Delta A_V^\textrm{int}$ also changes the $F_{\textrm{H}\alpha}/f_{\lambda \textrm{FUV}}$ of the most H$\alpha$-bright cluster observed, different models must be rejected for different $\Delta A_V^\textrm{int}$ due to the inability of producing clusters that are as H$\alpha$ bright as those we observed (see Fig. \ref{fig:maxfr}). For example, assuming $\Delta A_V^\textrm{int} = 0.1$ would allow all steep IMFs with $\alpha \la 3.1$. This is shown in Fig. \ref{fig:vAv} with hatches and dashed and dot-dashed lines that indicate the minimum of the predicted $\Sigma F_{\textrm{H}\alpha}/\Sigma f_{\lambda \textrm{FUV}}$ among our sets of steep (southwest to northeast hatches) and truncated (northwest to southeast hatches) IMF models, respectively, that are capable of producing clusters with as high a maximum $F_{\textrm{H}\alpha}/f_{\lambda \textrm{FUV}}$ as the maximum $F_{\textrm{H}\alpha}/f_{\lambda \textrm{FUV}}$ observed for a given $\Delta A_V^\textrm{int}$. The truncated IMF models cannot produce as low $\Sigma F_{\textrm{H}\alpha}/\Sigma f_{\lambda \textrm{FUV}}$ as is observed with any value of $\Delta A_V^\textrm{int}$, while the steep IMF models agree with observations with $0.10 < \Delta A_V^\textrm{int} \le 0.15$. Extrapolating the trend of the minimum $\Sigma F_{\textrm{H}\alpha}/\Sigma f_{\lambda \textrm{FUV}}$ of the steep models for $\Delta A_V^\textrm{int} > 0.15$ suggests models with $\alpha > 3.3$ would be able to reproduce the observed $\Sigma F_{\textrm{H}\alpha}/\Sigma f_{\lambda \textrm{FUV}}$ for $\Delta A_V^\textrm{int} > 0.15$. The lower panel in Fig. \ref{fig:vAv} shows how the observed $N_{\textrm{H}\alpha} / N_{\textrm{blue}}$ changes for the mass-selected catalog when assuming a Salpeter IMF and the mass-selected catalogs assuming steep IMFs as a function of $\Delta A_V^\textrm{int}$. We observed that $\Sigma F_{\textrm{H}\alpha}/\Sigma f_{\lambda \textrm{FUV}}$ and $N_{\textrm{H}\alpha} / N_{\textrm{blue}}$ are reconciled between observations and the steep IMF models for  $0.10 < \Delta A_V^\textrm{int} \le 0.15$ mag. Specifically, we found the best agreement with $\Delta A_V^\textrm{int} = 0.15$ mag and $\alpha = 3.2$.

We find  that the assumed extinction curve and the magnitude of extinction clearly have a large effect on the results of our analysis. However, no assumed value of internal extinction can match the observed ratios of our catalog to the Salpeter IMF model, while $0.10 < \Delta A_V^\textrm{int} \le 0.15$ mag creates a good match between our catalog and $\alpha > 3.1$ IMF models. Furthermore, this seems to suggest a higher internal extinction in the outer disk of M83 than the commonly assumed $A_V^\textrm{int} \approx 0$ in diffuse environments. Some recent studies have indeed shown that the extinction in diffuse environments may not always be negligible, such as \citet{junais2023dust} for a sample of LSB galaxies and Watkins et al. (in preparation) for the outer disk of M101.

\section{Discussion}
\label{sec:disc}

\subsection{Initial mass function in low-density environments}
Our results favor a steep $\alpha > 3.1$ IMF for the outer disk of M83, when assuming a constant SFH and $A_V^{\textrm{int}} \approx 0.15$ mag. The standard Salpeter IMF or simple truncated IMFs cannot explain our observations. The Salpeter IMF always overpredicts $\Sigma F_{\textrm{H}\alpha}/\Sigma f_{\lambda \textrm{FUV}}$, and the truncated IMFs cannot simultaneously reproduce the observed $\Sigma F_{\textrm{H}\alpha}/\Sigma f_{\lambda \textrm{FUV}}$ and the observed maximum $F_{\textrm{H}\alpha}/f_{\lambda \textrm{FUV}}$.

Our results apply to the integrated IMF over the outer disk of M83, which is equal to the intrinsic IMF of the star clusters only if the majority of the clusters are massive enough to fully sample the upper mass range of the intrinsic IMF. As suggested by \citet{pflamm2008imf}, this may not be the case in a low-SFR environment, such as the outer disk of M83, where instead the average cluster mass may be so low that a significant number of clusters are unable to generate O stars. However, our cluster count analysis is limited to massive clusters only, and as such, it should be free of any stochastic effects, yet we still measured lower $N_{\textrm{H}\alpha} / N_{\textrm{blue}}$ than predicted by the Salpeter IMF. Regardless, this scenario agrees better with our results than a hard density limit for massive star formation (e.g., \citealt{krumholz2008sf}), as it allows for the existence of individual clusters with high $F_{\textrm{H}\alpha}/f_{\lambda \textrm{FUV}}$ while explaining the more significant discrepancy of $\Sigma F_{\textrm{H}\alpha}/\Sigma f_{\lambda \textrm{FUV}}$ between our models and observations. Additionally, the steeper slope of the outer disk LFs indicates that low-mass clusters, which may not be able to generate O stars, most likely contribute to the total FUV flux of the outer disk relatively much more than to the total FUV flux of the inner disk. A similar scenario was proposed by \citet{meurer2009imf}, where a steeper or truncated IMF applies to loose or unbound clusters, which is more common in the outer disk due to the low-pressure environment. Regardless of the physics behind it, the integrated IMF of the M83 outer disk must be steep, such as the three-component IMF of \citet{elmegreen2004imf}, rather than truncated with a hard upper mass limit.

Constraining the exact shape of the intrinsic IMF over its entire mass range would require observations over more wavelengths and theoretical modeling, which is beyond the scope of this work. However, our analysis shows that the integrated IMF in the outer disk of M83 is deficient in high-mass stars, or top-light, compared to the standard Salpeter IMF or to the integrated IMF in the inner disk of M83. Our results complement the evidence for a top-light IMF in LSB galaxies found in the surveys of integrated galaxy properties by \citet{hoversten2008imf}, \citet{meurer2009imf}, and \citet{lee2009hafuv}.

\subsection{Comparison to previous studies of the M83 outer disk}
Even though we share the same target and the same wavelengths, our H$\alpha$-to-FUV flux ratio measurements differ from those of \citetalias{koda2012imf}, and consequently, our conclusions regarding the universality of the IMF differ as well. The main argument for the Salpeter IMF in the M83 outer disk of \citetalias{koda2012imf} was that the observed counts of H$\alpha$-bright and blue clusters matched a \textsc{Starburst99} model using the standard Salpeter IMF. We found $N_{\textrm{H}\alpha} / N_{\textrm{blue}}$ significantly lower than that predicted by the Salpeter IMF, and we also note that the predicted $N_{\textrm{H}\alpha} / N_{\textrm{blue}}$ is practically invariant of the IMF used in the models, as long as the IMF is capable of producing O stars. An internal extinction value of $A_V^{\textrm{cluster}} = 0.1$ was also adopted by \citetalias{koda2012imf} for the M83 outer disk star clusters, supported by the agreement in color-color space it created between their observations and model. This observation agrees well with our result that $0.10 < A_V^{\textrm{cluster}} < 0.15$ creates the best match in $F_{\textrm{H}\alpha}/f_{\lambda \textrm{FUV}}$ and $N_{\textrm{H}\alpha} / N_{\textrm{blue}}$ between our observations and models.

\citet{bruzzese2020imf} also studied the IMF in the outer disk of M83. They used \emph{Hubble Space Telescope} Advanced Camera for Surveys Wide Field Camera (ACS/WFC) imaging and H$\alpha$ data from the Cerro Tololo Inter-American Observatory (CTIO) to analyze the resolved stellar populations in four fields within the M83 outer disk. They found that a truncated IMF with $M_{\textrm{u}} = 25 M_{\odot}$ fits best with their observations. Their best-fit IMF for just the ACS/WFC imaging has $\alpha=2.35$, while when they use their H$\alpha$ data as an additional constraint, the best-fit IMF has $\alpha=1.95$. Though our results agree in general with \citet{bruzzese2020imf} that the IMF in the M83 outer disk is deficient in high-mass stars, our analysis does not support a similar hard truncation of the IMF. Each of the ACS/WFC fields used by \citet{bruzzese2020imf} covers an area of $202\arcsec \times 202\arcsec$, this is only a small fraction of the outer disk of M83 and approximately $1 \%$ of our coverage. It may be that due to the rarity of massive O stars in the M83 outer disk, the areas studied by \citet{bruzzese2020imf} are missing them simply because of stochastic effects or due to local variations of the IMF. In fact, we found no H$\alpha$-bright clusters within their field W4, and we found clusters requiring stars more massive than $30 M_\odot$ only in their field W3, indicating that the clusters containing the most massive stars in the outer disk are indeed located outside the regions studied by \citet{bruzzese2020imf}. Additionally, the $\alpha=1.95$ IMF suggested by their H$\alpha$ data would produce clusters that are more H$\alpha$ bright than an IMF with a Salpeter slope ($\alpha = 2.35$) and the same truncation mass ($M_{\textrm{u}} = 25 M_{\odot}$), possibly fitting better to our observations than the truncated IMFs we tested. Overall, \citet{bruzzese2020imf} extend the evidence against the standard Salpeter IMF in the outer disk of M83 to optical wavelengths, strongly supporting the notion that IMF is not universal in low-density environments. In conclusion, while there are discrepancies between our results and those of \citetalias{koda2012imf}, our new results confirm the top-light IMF in the M83 outer disk reported by \citet{bruzzese2020imf}, and these results extend the evidence for the dependence of the integrated IMF to environmental parameters.

\subsection{Catalog limitations}
Our catalog is approximately $80\%$ complete at our adopted detection limit of FUV $= 23.8$ mag. Clusters of mass $2 \times 10^3 M_\odot$ following a Salpeter IMF as well as clusters of mass $\sim 10^4 M_\odot$ following a steep IMF will fall below this magnitude after reaching 100 Myr in age. This means that we cannot detect low-mass clusters ($< 10^3 M_\odot$) older than a few tens of million years. Our catalog is therefore biased toward younger and more massive clusters. However, this bias cannot explain our main result, that is, that our observations are incompatible with a Salpeter IMF in the M83 outer disk. As old low-mass clusters are redder and lack in H$\alpha$ emission compared to the clusters in our catalog, missing a significant fraction of them would bias our observed $N_{\textrm{H}\alpha} / N_{\textrm{blue}}$ and $\Sigma F_{\textrm{H}\alpha}/ \Sigma f_{\lambda \textrm{FUV}}$ toward larger values, meaning that the true $N_{\textrm{H}\alpha} / N_{\textrm{blue}}$ and $\Sigma F_{\textrm{H}\alpha}/ \Sigma f_{\lambda \textrm{FUV}}$ would differ from the Salpeter IMF model predictions even more than our measured values. In other words, the bias in our catalog cannot explain the differences we observe in the M83 outer disk compared to predictions from the Salpeter IMF models.

The coarse \emph{GALEX} resolution imposes another limitation on our catalog. Multiple small clusters within a distance corresponding to the \emph{GALEX} PSF FWHM ($\sim$100 pc) from each other may be detected as a single larger cluster. However, this sort of blending of sources must be very rare in the M83 outer disk due to the sparse distribution of clusters therein. Furthermore, it would not affect integrated quantities such as $\Sigma F_{\textrm{H}\alpha}/ \Sigma f_{\lambda \textrm{FUV}}$, and its effect on $N_{\textrm{H}\alpha} / N_{\textrm{blue}}$ is unclear, as both blue and H$\alpha$-bright clusters would suffer from this blending.

\subsection{Star formation history}
Our analysis relies on the assumption of a constant cluster formation rate over the entire outer disk. If instead the cluster formation rate was greater at some point in time more than 10 Myr ago but less than 1 Gyr ago, we would expect to find more old clusters that are not H$\alpha$ bright but may be blue and contribute to the total FUV emission of the outer disk. This could explain why the model with the standard Salpeter IMF overpredicts $N_{\textrm{H}\alpha} / N_{\textrm{blue}}$ and $F_{\textrm{H}\alpha}/f_{\lambda \textrm{FUV}}$.

This kind of variation in the global SFR over the M83 outer disk could be a result of bursty star formation. In dwarf galaxies, \citet{hoversten2008imf} found that the SFH is more likely to be bursty than smoothly varying. In diffuse environments, such as dwarf galaxies or the outer disk, individual cloud collapse star-forming events may be rare, but they may trigger further star formation through gas hydrodynamics or massive star supernovae. In this scenario, the duration of the global star-forming burst is limited by the speed at which the phenomenon propagates through the outer disk. Assuming that a disturbance that triggers star formation propagates at a speed equal to the H{~\sc i} velocity dispersion in the outer disk ($\sim 15$ km s$^{-1}$; \citealt{heald2016hi}), it would take it more than 3 Gyr for it to travel the $\sim50$ kpc breadth of the M83 outer disk. This means that such a spontaneously triggered star-forming event in the past should have left azimuthal or radial gradients in $F_{\textrm{H}\alpha}/f_{\lambda \textrm{FUV}}$, as clusters become faint in FUV near the origin of the event before it has reached the far edges of the outer disk. As we found no azimuthal or radial variation in $F_{\textrm{H}\alpha}/f_{\lambda \textrm{FUV}}$ in the outer disk of M83, this type of bursty SFH can be ruled out.

Another phenomenon that could be responsible for temporal variations in the outer disk SFR is galaxy interactions.  Within the Cen A Group, M83 is the central galaxy of its own subgroup, and three dwarf companions (IC~4316, NGC~5264, and UGCA~365) have been found within a 100 kpc projected distance of its center \citep{koribalski2018hi}. As such, it is likely that M83 has experienced close encounters and possible mergers with smaller companions in its past. The distorted, one-armed appearance of the H{~\sc i} disk of M83 is also suggestive of this. Assuming typical relative speeds of a few hundred kilometers per second (such as that of the Sagittarius Dwarf of the Milky Way, \citealt{law2010sgr}), an infalling dwarf galaxy could cross the outer disk of M83 within a few hundred million years, triggering star formation across the outer disk. An event of this kind could have occurred between 10 Myr and 1 Gyr ago, triggering enhanced cluster formation and leaving behind clusters that still contribute to the total FUV flux of the outer disk but are no longer bright in H$\alpha$ nor blue in UV. However, it is unclear how significant the increase of SFR triggered by this kind of event would be and if the initial starburst would occur in a large enough area so that the slower secular propagation would not leave behind noticeable age gradients.

\subsection{Evolved stars and scattered light}
The \textsc{Starburst99} code specializes in modeling young stellar populations, and its treatment of evolved stars, such as white dwarfs, is simplistic, and for some cases, such as post-asymptotic giant branch stars, is missing completely. Still, these very hot evolved stars may be a significant source of FUV emission in galaxies (e.g., \citealt{hills1972ism, flores-fajardo2011holmes, rautio2022}). It could be that some of the discrepancy in $F_{\textrm{H}\alpha}/f_{\lambda \textrm{FUV}}$ between our Salpeter IMF model and the observations comes from our model missing this population of very old stars. Additionally, some evolved stars could also be ejected from the inner disk to the outer disk during their long lives, increasing the number of old stars compared to any in situ models. However, it is known that old stars outside globular clusters form a very diffuse and even background population, as their unbound birth clusters disperse. They would appear in our data as a disk of diffuse FUV light on the scale of the galaxy, and our background subtraction described in Sect. \ref{sec:galex} would subtract them out. Therefore, the excess of FUV emission we measure cannot come from any diffuse evolved population of stars.

Another potential source of FUV emission in the outer disk is starlight originating from the inner disk but scattered to the line of sight by dust in the outer disk. Dust scattering starlight can be a significant source of emission, specifically for edge-on galaxies and in UV (e.g., \citealt{jo2018fuv}). However, the outer disk of M83 extends tens of kiloparsecs outside the inner disk, compared to the $\sim 1$ kpc scale heights of extraplanar dust, so any FUV emission originating from the inner disk must cross much larger distances through much more optically thick media as it propagates along the midplane to reach any dust in the outer disk, compared to the FUV emission scattered by extraplanar dust. As such, it seems unlikely that a significant amount of outer disk FUV emission could be reflected starlight originating from the inner disk. Additionally, any starlight scattered by diffuse dust would appear as a diffuse emission on the scale of the galaxy, and again our background subtraction would subtract it out.

\section{Summary and conclusions}
\label{sec:sum}
In this work, we used archival \emph{GALEX} FUV and NUV data and new deep OmegaCAM H$\alpha$ imaging to investigate the IMF in the outer disk of M83. To do this, we constructed a catalog of FUV-selected objects in the outer disk and compared the counts of H$\alpha$-bright and blue clusters ($N_{\textrm{H}\alpha} / N_{\textrm{blue}}$), maximum H$\alpha$-to-FUV flux ratios of the clusters ($F_{\textrm{H}\alpha}/f_{\lambda \textrm{FUV}}$), and the H$\alpha$-to-FUV total flux ratio ($\Sigma F_{\textrm{H}\alpha}/\Sigma f_{\lambda \textrm{FUV}}$) over the outer disk to predictions obtained from \textsc{Starburst99} stellar population synthesis models.

We computed two sets of models by varying the truncation mass ($20 M_\odot < M_{\textrm{u}} < 90 M_\odot$) and IMF slope ($2.5 < \alpha < 3.3$) separately, and we compared these to a model with the standard Salpeter IMF. When taking into account only $A_V^{\textrm{MW}}$ and $A_V^{\textrm{int}}$ calculated from H{~\sc i} data with no additional internal extinction, none of our models could simultaneously reproduce all of the three observational constraints that we considered, namely the $N_{\textrm{H}\alpha} / N_{\textrm{blue}}$, the maximum $F_{\textrm{H}\alpha}/f_{\lambda \textrm{FUV}}$, and the $\Sigma F_{\textrm{H}\alpha}/\Sigma f_{\lambda \textrm{FUV}}$. The steep models underestimate $N_{\textrm{H}\alpha} / N_{\textrm{blue}}$, and the truncated models cannot produce clusters as H$\alpha$ bright as the maximum observed, while the Salpeter IMF overestimates the $\Sigma F_{\textrm{H}\alpha}/\Sigma f_{\lambda \textrm{FUV}}$ by a factor of approximately three.

Previous spectroscopic studies have reported a large scatter in the total extinctions of the M83 outer disk stellar clusters, and we show that assumed internal extinction has a large impact on $N_{\textrm{H}\alpha} / N_{\textrm{blue}}$ and $\Sigma F_{\textrm{H}\alpha}/\Sigma f_{\lambda \textrm{FUV}}$. By repeating our analysis with different values of additional assumed internal extinction, we found that with $0.10 < \Delta A_V^\textrm{int} < 0.15$, a steep IMF model with $\alpha > 3.1$ can reproduce our observations well. Dedicated extinction measurements in the outer disk of M83 are required to confirm this result. We also found that in the inner disk, $\Sigma F_{\textrm{H}\alpha}/\Sigma f_{\lambda \textrm{FUV}}$ is much higher than in the outer disk and closer to the Salpeter IMF value.

Our results support a  steep, non-universal IMF in the diffuse outskirts of M83. While the IMF must be top-light, a simple truncation must be ruled out due to the existence of clusters that are too H$\alpha$ bright in the outer disk. A speculation on the exact shape of the intrinsic IMF in this diffuse environment and its relation to the physical parameters of its environment is beyond the scope of this work. The true IMF in the M83 outer disk may be both slightly truncated and steeper than the Salpeter IMF, and it does not  even need to follow a single power-law index, as our simplistic models do. Indeed, a three-part IMF where potentially differing formation processes for low-, intermediate-, and high-mass stars are taken into account, such as the one suggested by \citet{elmegreen2004imf}, may better reflect reality. Further work on galaxy outskirts and other low-density environments are required to confirm the non-universality of the IMF and further constrain it. The non-universality of the IMF has significant implications on recipes for determining SFRs and SFHs from photometric properties of stellar populations, and if confirmed, these must be adjusted accordingly in low-density environments.

\begin{acknowledgements}
  We thank the referee for most useful comments that helped to quantify the detection limit and completeness of our cluster sample.
  This research is based on observations collected at the European Southern Observatory under ESO programme 106.217Q.001; and on observations made with the Galaxy Evolution Explorer, obtained from the MAST data archive at the Space Telescope Science Institute, which is operated by the Association of Universities for Research in Astronomy, Inc., under NASA contract NAS 5–26555.
  RR acknowledges funding from the Technology and Natural Sciences Doctoral Program (TNS-DP) of the University of Oulu.
  HS and AW acknowledge funding from the Academy of Finland grant n:o 297738.
  AW additionally acknowledges support from the STFC [grant numbers ST/S00615X/1 and ST/X001318/1].
  AV acknowledges funding from the Academy of Finland grant n:o 347089.
  SC acknowledges funding from the State Research Agency (AEI-MCINN) of the Spanish Ministry of Science and Innovation under the grant  “Thick discs, relics of the infancy of galaxies" with reference PID2020-113213GA-I00.
 We acknowledge the usage of the NASA/IPAC Extragalactic Database (NED) which is operated by the Jet Propulsion Laboratory, California Institute of Technology, under contract with the National Aeronautics and Space Administration.
 This research made use of python (\url{http://www.python.org}), IDL (\url{http://www.harrisgeospatial.com}), SciPy \citep{2020SciPy}, NumPy \citep{2020NumPy}, Matplotlib \citep{matplotlib}, NoiseChisel \citep{gnuastro,noisechisel}, and Astropy, a community-developed core Python package for Astronomy. This work was partly done using GNU Astronomy Utilities (Gnuastro, ascl.net/1801.009) version 0.19. Work on Gnuastro has been funded by the Japanese Ministry of Education, Culture, Sports, Science, and Technology (MEXT) scholarship and its Grant-in-Aid for Scientific Research (21244012, 24253003), the European Research Council (ERC) advanced grant 339659-MUSICOS, the Spanish Ministry of Economy and Competitiveness (MINECO, grant number AYA2016-76219-P) and the NextGenerationEU grant through the Recovery and Resilience Facility project ICTS-MRR-2021-03-CEFCA.
 We also acknowledge Joachim Janz for participating in drafting the proposal for the VST observations.
\end{acknowledgements}

\bibliographystyle{aa}
\bibliography{ref}{}

\end{document}